\documentclass[english,a4paper,eqno,11pt]{article}

\usepackage[english]{babel}
\usepackage{graphicx,epsfig,epic} 
\usepackage{pifont}      

\usepackage{pict2e}

\def\cal{\mathcal}

\def\esp{\vskip .6cm}
\def\pesp{\vskip .3cm}

\def\ni{\noindent}
\def\di{\displaystyle}

\newtheorem{defn}{Definition}

\newtheorem{prop}{Proposition}
\newtheorem{prp}{Properties}
\newtheorem{cor}{Corollary}

\newtheorem{rem}{Comment}
\newtheorem{rmak}{Remark}

\begin{document}

\title{Special extension of the  relativity of moving bodies: \\ A Case study}
\author{\small Fay\c cal Ben Adda\\ \small Community College of Qatar\\
\small{Email: f\_benaddafr@yahoo.fr}}

\maketitle

\begin{abstract}
The incorporation of an adequate discrete expansion to the formalism of the special relativity that does not allow gravitational acceleration unravels unexplored phenomena. This extension takes into account consequences of a small variation of the limiting velocity for a large scale of time together with the space-time expansion. It determines a natural boundary between physical systems with high and low kinetic energy, provides a new insight for Earth revolution since its formation, indicates a new factor that maintains the earlier remnant heat inside Earth, indicates that the self-sustaining nuclear fission chain reaction that generates Earth major heat is increasing together with the space-time expansion, provides a new lead for the problem of galaxies missing mass, shows how time is entangled with space and expansion and leads to a new interpretation of its characteristics.



{\footnotesize AMS: 38A05,  81V35, 83F05,  26D07,  26A12,  26A24, 40B05}

\end{abstract}

\section{Introduction}
\label{intro}

The special relativity theory is a theory of measurements of events together with their linear transformations in inertial reference frames, that extends Newtonian mechanics for physical systems with relatively high speed. This theory was proposed by Albert Einstein in 1905 (\cite{EI3}), after substantial contributions of H. Lorentz and H. Poincar\'e (\cite{HP1},\cite{HP2}). This theory led to the discovery of nuclear energy and unlocked the secret of stars. In 1915 Albert Einstein proposed a new theory for reference frames that undergo gravitational acceleration, called general relativity (without being a generalization of the special relativity) in which  the special relativity theory appears to be one particular solution only valid for a flat space-time without acceleration. The general relativity theory suggests that gravity, as well as motion, can affect the intervals of space and time, and leads to the space warps, the Big Bang, and the discovery of black holes in our universe.

Despite the fact that special relativity is a particular solution, it was extended to encompass physical system with higher speed and used in particle physics (in quantum field theory \cite{ManCH}, \cite{PCHRO},\cite{Wein1},\cite{Wein2}). Within this case study we propose a new extension of the special relativity theory that might reveal unquestionable phenomena unexplored before. We propose to incorporate an adequate discrete expansion of the space to the formalism of the special relativity theory, that does not allow the inertial reference frame to undergo gravitational acceleration. This extension allows to reevaluate the measurement of events and their linear transformations within inertial reference frames, which leads to rewrite and interpret the relativistic laws of physics in the light of the special expansion, and explore applications.

The plan of this framework is as follow: In section 2, we introduce a discrete expansion and provide an adequate space-time interval. In section 3, the Lorentz transformations equations are presented with their consequences. In section 4 the composition law of velocity are introduced. In section 5, momentum and relative mass are formulated. In section 6, kinetic energy, total energy and  energy variation together with the space-time expansion are obtained and discussed. In section 7, time together with the space-time discrete expansion is explored. In section 8, we introduce some applications

\section{Space-time interval with discrete expansion}

\subsection{Discrete expansion}

We start the quantification of the space-time expansion from the primordial universe around 380,000 years after the Big Bang, when the expansion changes the universe from opaque universe to transparent universe that makes light (photon) free to travel and then the cosmic microwave background (CMB) to appear. The seeds of the cosmic structure appear to be traced by a tiny fluctuation recorded in the CMB that allocates positions of matter distribution in the first earlier observed universe called the primordial universe. We propose to subdivide the period of time of the expansion of the space-time starting from the primordial universe to nowadays into n periods of large scale of time called Steps, where the Step 0 represents the primordial universe and the Step n represents the present space-time, and where the space-time expands from the Step 0 to the Step n following a discrete space-time expansion quantified by $n\geq0$. The bigger the subdivision n of the expansion period is, the smaller the scale of time is. We call  $\mathcal E$ this expanding space-time characterized as follow:
\begin{defn}\label{para}
Let $L_0\not=0$ be the distance between any two separated events in $\mathcal E$ at the Step 0, where  $L_n$ is their distance at the Step n, and $L_{n+1}$ is their distance at the Step n+1  in the expanding space-time $\mathcal E$. The distance $L_{n+1}$ at the Step (n+1) is defined by $L_{n+1}=a_{n}L_{n}$ for all $n\geq0$, where $(a_n)_{\geq0}$ is a numerical sequence such that $a_0=1$, $a_n>1$ $\forall n\geq1$, and where the product $\prod_{i=0}^na_i$ is a convergent expanding parameter.
\end{defn}
 Some applications of the discrete expansion can be found in (\cite{BF2},\cite{BF3},\cite{BFP},\cite{BF0},\cite{BF4}). The discrete expansion verifies:
\begin{cor}
Using the above discrete expansion of the space-time $\mathcal E$, if the distance between two events at the Step n is equal to $L_n$ for all $n\geq0$ then

i) for all $n\geq0$, $L_{n}=L_{0}(\prod_{i=0}^na_i)$,

ii) for all $n\geq0$, $L_{n+1}>L_n$.
\end{cor}

{\it Proof}: i) Since $L_{n+1}=a_{n+1}L_{n}$ for all $n\geq0$, then $L_{1}=a_{1}L_{0}$ and  $L_{2}=a_{2}L_{1}=a_{2}a_{1}L_{0}$. Which gives by iteration $L_{n+1}=a_{n+1}L_{n}=a_{n+1}a_{n}a_{n-1}\ldots a_1 L_0$. Since $a_0=1$, then
\begin{equation}
L_{n+1}=l_0(\prod_{i=1}^{n+1}a_i)=l_0(a_0\prod_{i=1}^{n+1}a_i)=l_0(\prod_{i=0}^{n+1}a_i),
\end{equation}
then for all $n\geq0$, $L_{n}=L_{0}(\prod_{i=0}^na_i)$.

ii) For all $n\geq0$
\begin{equation}
{L_{n+1}\over L_n}={L_{0}(\prod_{i=0}^{n+1}a_i)\over L_{0}(\prod_{i=0}^{n}a_i)}=a_{n+1}>1,
\end{equation}
that is to say $L_{n+1}>L_n$ for all $n\geq0$ which gives ii).

\subsection{The space-time interval}

Using the discrete quantification to record events starting from the primordial universe (Step 0) to present time (Step n), consider un event $A_n$ for all $n\geq0$ observed at each Step n in an inertial reference frame $\mathcal R_n$ as follow:

\textbf{Step 0}: The Step 0 represents the primordial space-time $\mathcal E$ and is defined as the set of events together with the notion of space-time interval $S_0^2$ given as follow : if an event $A_0$ is recorded in an inertial reference frame $\mathcal R_0$ by four coordinates $x$, $y$, $z$  and $t$, where $(x,y,z)$ are the spatial coordinates of the event, and $t$ fixes the instant when this event is recorded, then the squared space-time interval $\mathcal S_0^2$ of the event is given as function of these quantities through the generalization of Pythagoras theorem
\begin{equation}
\mathcal S_0^2=(ct)^2-a^2_0\Big( x^2+ y^2+ z^2\Big)
\end{equation}
where $c$ is the maximum speed of light propagation recorded by an observer in $\mathcal R_0$ when the universe becomes transparent and light was released to travel for the first time after the Big Bang, and where $a_0=1$.

\textbf{Step n}: The Step n represents the space-time at present time after n steps of discrete expansion for all $n\geq0$, in which the event $A_n$ is recorded in the inertial reference frame $\mathcal R_n$ by four coordinates $x$, $y$, $z$ and $t$, where $(x,y,z)$ are the spatial coordinates of the event and $t$ fixes the instant when the event $A_n$ is recorded by an observer. The squared space-time interval $\mathcal S_n^2$ is then given as function of the event coordinates through the generalization of Pythagoras theorem
\begin{equation}
\mathcal S_n^2=(ct)^2-\Big(\prod_{i=0}^{n}a_i^2\Big)(x^2+y^2+z^2)
 \end{equation}
where $c$ is the maximum speed of light propagation recorded by an observer from the inertial reference frame $\mathcal R_0$ in the primordial universe.

\begin{defn}\label{Md}
Using the discrete quantification of the space-time expansion $\mathcal E$, the squared space-time interval at the Step n is defined for all $n\geq0$  by
\begin{equation}\label{M1}
\mathcal S^2_n=(c t)^2-\Big ( \prod_{i=0}^na_i\Big)^2( x^2+ y^2+ z^2),
\end{equation}
where $(x,y,z)$ are the spatial coordinates of the event recorded by an observer relative to an inertial reference frame $\mathcal R_n$ at the Step n, and where $t$ fixes the instant when this event is recorded relative to the frame $\mathcal R_n$ at the Step n, $c$ is the maximum speed of light propagation recorded by an observer from the inertial reference frame $\mathcal R_0$ in the primordial universe (the Step 0).
\end{defn}
\begin{rmak}\label{rmak}
1) If the event $A_n$ for all $n\geq0$ is observed at each Step n in an inertial reference frame $\mathcal R_n$ to occur at the same space position $(x,y,z)$, the space-time interval (\ref{M1}) conveys how the space time is stretched in all direction with the sequence $(a_n)_{n\geq0}$ of expanding parameters introduced in Definition \ref{para}.

2) The parameter $n$ in the equality (\ref{M1}) represents the subdivision for the period of time between the primordial universe to present time. If the parameter $n$ is a small number then the period of time between one step to another is large, and if the parameter $n$ is a big number then the period of time between one step to another is small.
\end{rmak}

\section{Lorentz transformations equations}

To rewrite the Lorentz transformations using the space-time $\mathcal E$ associated with a discrete quantified expansion in which the space-time interval is given by (\ref{M1}) such that at each Step n for all $n\geq0$ the equations of Newtonian mechanics hold good, and at small scales as well as at the sub-galactic scales the fundamental forces of nature remain valid we introduce the following:

Consider an event $A$ in the space-time $\mathcal E$ at the Step n for all $n\geq0$ and consider for all $n\geq0$ two inertial references frame $\mathcal R_n$ and $\mathcal R_n'$, where the frame $\mathcal R_n'$ has a velocity $\overrightarrow{v}$ relative to the frame $\mathcal R_n$ parallel to their common positive $x$-axis. An observer in the frame $\mathcal R_n$ reports for all $n\geq0$ the space-time coordinates $x$, $y$, $z$, $t$ for the event $A$, where the space-time interval relative to the frame $\mathcal R_n$ is given by (\ref{M1}) and satisfies for all $n\geq0$
\begin{equation}\label{T01}
\mathcal S^2_n=(c t)^2-\Big(\prod_{i=0}^na_i\Big)^2( x^2+y^2+z^2),
\end{equation}
and an observer in $\mathcal R_n'$ reports $x'$, $y'$, $z'$, $t'$ for the same event $A$, where the space-time interval relative to the frame $\mathcal R_n'$ is given by (\ref{M1}) and satisfies for all $n\geq0$
\begin{equation}\label{T02}
\mathcal {S'}^2_n=(c t')^2-\Big ( \prod_{i=0}^na_i\Big)^2( x'^2+ y'^2+ z'^2),
\end{equation}
assuming for $t=t'=0$ the origins of $\mathcal R'_n$ and $\mathcal R_n$ coincide and that the coordinates $(x,y,z,t)$ and $(x',y',z',t')$ are related by the linear transformation
\begin{equation}\label{E01}
T_n:\left\{
  \begin{array}{ll}
    x'= & a_{11}x+a_{14}t \\
    y'= & a_{22}\ y \\
    z'= & a_{33}\ z\\
    t'= & a_{41}\ x+a_{44}\ t
  \end{array}
\right.
\end{equation}

In general for a linear transformation of the Cartesian frames at the Step n for all $n\geq0$, we have
\begin{equation}\label{gen}
\left\{
  \begin{array}{ll}
    x'= & a_{11}x+a_{12}y+a_{13}z+a_{14}t \\
    y'= & a_{21}x+a_{22}y+a_{23}z+a_{24}t \\
    z'= & a_{31}x+a_{32}y+a_{33}z+a_{34}t \\
    t'= & a_{41}x+a_{42}y+a_{43}z+a_{44}t
  \end{array}
\right.
\end{equation}
where the coefficients $a_{ij}$ depend on the movement. The general properties of homogeneity and isotropy of the space with the choice that the frame $\mathcal R'_n$ moves in the positive x direction of the frame $\mathcal R_n$ with uniform velocity $v$, such that the corresponding axes of $\mathcal R_n$ and $\mathcal R'_n$ remain parallel for all $n\geq0$ throughout the motion having coincided at $t=t'=0$ at the Step n for all $n\geq 0$, will reduce the equations (\ref{gen}), similar to the classical approach, to the transformation  $T_n$ given by (\ref{E01}).

\begin{prop}\label{Ltransf}
Assuming that the frame $\mathcal R_n'$ has a constant velocity with magnitude $\di v<{c\over\prod_{i=0}^n a_i}$ at the Step n for all $n\geq0$ relative to the frame $\mathcal R_n$ introduced above, where $\prod_{i=0}^n a_i$ is the convergent expanding parameter introduced in Definition \ref{para}. If the linear transformation (\ref{E01}) is a solution of the equation  $\mathcal {S}^2_n=\mathcal {S'}^2_n$ for a strictly positive coefficient $a_{11}$, and $a_{44}$, then the linear transformation (\ref{E01}) for all $n\geq 0$ is equal to
\begin{equation}\label{E4g}
T_n:\left\{
  \begin{array}{ll}
    x'= & \di{x-vt\over\sqrt{1-(\prod_{i=0}^n a_i^2){v^2\over c^2}}} \\
    y'= & y \\
    z'= & z\\
    t'= & \di{t-(\prod_{i=0}^n a_i)^2{vx\over c^2}\over\sqrt{1-(\prod_{i=0}^n a_i^2){v^2\over c^2}}}
  \end{array}
\right.
\end{equation}
where $c$ is the constant speed of light recorded at the primordial universe (Step 0), and $\mathcal {S}^2_n$ and $\mathcal {S'}^2_n$ are given by (\ref{T01}) and (\ref{T02}).
\end{prop}

{\it Proof}:
Starting from equation  $\mathcal S^2_n={\mathcal S'}^2_n$ for all $n\geq0$, we have
\begin{equation}\label{E2}
(ct)^2-\di\Big(\prod_{i=0}^n a_i\Big)^2(x^2+y^2+z^2)=(ct')^2-\Big(\prod_{i=0}^n a_i\Big)^2(x'^2+y'^2+z'^2).
\end{equation}

If the linear transformation (\ref{E01}) is a solution of the equality (\ref{E2}), then the substitution of (\ref{E01}) in (\ref{E2}) gives
$$(ct)^2-\prod_{i=0}^n a_i^2(x^2+y^2+z^2)=c^2\prod_{i=0}^n a_i^2(a_{41}x+a_{44}t)^2-\prod_{i=0}^n a_i^2\Big((a_{11}x+a_{14}t)^2+(a_{22}y)^2+(a_{33}z)^2
\Big)$$
which yields, using polynomial identification, the following system
\begin{equation}\label{E3}
\left\{
  \begin{array}{ll}
    \prod_{i=0}^n a_i^2\ = & (\prod_{i=0}^n a_i^2)\ a_{11}^2-c^2a_{41}^2 \\
    \quad 1 \qquad= & a_{22}^2 \\
   \quad 1 \qquad = & a_{33}^2\\
    -c^2 \qquad = &(\prod_{i=0}^n a_i^2)\ a_{14}^2-c^2a_{44}^2\\
   \quad 0 \qquad = & -(\prod_{i=0}^n a_i^2)\ a_{11}a_{14}+c^2a_{41}a_{44},
  \end{array}
\right.
\end{equation}
following the classical approach, the last equation in (\ref{E3}) gives
 \begin{equation}\label{int1}
({c^2\over\prod_{i=0}^n a_i^2}){a_{41}\over a_{11}}={a_{14}\over a_{44}}=v
 \end{equation}
(since for $x=0$ in (\ref{E01}) gives $\di x'={a_{14}\over a_{44}}t'$ and we know that  $x'=vt'$, then $\di{a_{14}\over a_{44}}=v$).

From (\ref{int1}) we can deduce
\begin{equation}\label{a41}
a_{41}=({\prod_{i=0}^n a_i^2\over c^2})\ v\ a_{11}
\end{equation}
 and
 \begin{equation}\label{a14}
a_{14}= v\ a_{44}.
\end{equation}

Since $a_{11}>0$, the substitution of (\ref{a41}) in the first equality of (\ref{E3}) gives for $\di v<{c\over\prod_{i=0}^n a_i}$
\begin{equation}\label{a11}
a_{11}={1\over\sqrt{1-{\prod_{i=0}^n a_i^2{v^2\over c^2}}}},
\end{equation}
and the substitution of (\ref{a14}) in the fourth equality of (\ref{E3}), for $a_{44}>0$ and $\di v<{c\over\prod_{i=0}^n a_i}$ gives
\begin{equation}\label{a44}
a_{44}={1\over\sqrt{1-{\prod_{i=0}^n a_i^2{v^2\over c^2}}}},
\end{equation}
and the substitution of (\ref{a11}) in (\ref{a41}) and (\ref{a44}) in (\ref{a14}) gives
\begin{equation}\label{a410}
a_{41}=\prod_{i=0}^n a_i^2{v\over c^2}{1\over\sqrt{1-{\prod_{i=0}^n a_i^2{v^2\over c^2}}}},\qquad\hbox{and}\qquad a_{14}={v\over\sqrt{1-{\prod_{i=0}^n a_i^2{v^2\over c^2}}}},
\end{equation}
finally the substitution of  (\ref{a410}), (\ref{a44}), and (\ref{a11}) in (\ref{E01}) leads to the Lorentz transformation (\ref{E4g}).
\esp
Since the velocity $\overrightarrow{v}$ of the inertial frame $\mathcal R'_n$ relative to the frame $\mathcal R_n$ is chosen to be parallel to their common x-axis, and if we denote ($x_n$, $y_n$, $z_n$, $t_n$) as the space-time coordinates in $\mathcal R_n$ and ($x'_n$, $y'_n$, $z'_n$, $t'_n$) as the space-time coordinates in $\mathcal R'_n$ for all $n\geq0$, the transformations (\ref{E4g}) can be reduced using the differences between coordinates for a pair of events measured by observers in $\mathcal R_n$ as well as in $\mathcal R'_n$ by
\begin{equation}\label{r2E}
T_n:\left\{
  \begin{array}{ll}
    \Delta x'_n=\gamma_n({\Delta x_n- v\ \Delta t_n}) &  \\
    \Delta t'_n= \gamma_n({\Delta t_n-{v\ \Delta x_n\over c_n^2}})&
  \end{array}
\right.
T_n^{-1}:\left\{
  \begin{array}{ll}
    \Delta x_n=\gamma_n({\Delta x'_n+v\ \Delta t'_n}) &  \\
    \Delta t_n=\gamma_n({\Delta t'_n+{v\ \Delta x'_n\over c_n^2}}) &
  \end{array}
\right.
\end{equation}
where \begin{equation}\label{gn}
c_n={c\over\prod_{i=0}^n a_i}, \qquad\hbox{and}\qquad \gamma_n={1\over\sqrt{1-{v^2\over c_n^2}}}.
\end{equation}
\subsection{Properties from Lorentz transformations equations}
 From the Lorentz transformations (\ref{r2E}) we have the following:
\subsubsection{Simultaneity}
Two simultaneous events at the Step n ($\Delta t_n=0$) in the reference frame $\mathcal R_n$ for all $n\geq0$ are not simultaneous in $\mathcal R'_n$ since (\ref{r2E}) gives $\Delta t'_n=-{v\ \Delta x_n\over c_n^2}$, the simultaneity is affected by the magnitude $v$ of the reference frame as well as the expanding parameter of the space-time since $c_n$ is given by (\ref{gn}) .
\subsubsection{Time dilation}\label{tdilation}
If two events occurs at the same place in the reference frame $\mathcal R'_n$ for all $n\geq0$ but at different times, that is to say $\Delta x'_n=0$ for all $n\geq0$ and time $\Delta t'_n=\Delta\tau\neq0$, then equation (\ref{r2E}) gives the time dilation $\Delta t_n=\gamma_n \Delta\tau$.

a) at each Step n, the time dilation is affected by the magnitude of the relative velocity  of the reference frame since for $v_2<v_1<c_n$, we have $\gamma_{n}(v_2)<\gamma_{n}(v_1)$.

b) Moreover, the time dilation is also affected by the space-time expanding parameter since at the Step n+1, and under the same conditions equation (\ref{r2E}) gives also the time dilation $\Delta t_{n+1}=\gamma_{n+1} \Delta \tau$, and since $\gamma_{n+1}(v)>\gamma_{n}(v)$ for $v<c_n$, then $\Delta t_{n+1}>\Delta t_{n}$.
\subsubsection{Length contraction}
For all $n\geq 0$, consider an object at rest in the inertial reference frame  $\mathcal R'_n$ with a proper length parallel to the $x_n$-axis given by $l_0=\Delta x'_n$. If the length of the object is measured  by another observer in the reference frame $\mathcal R_n$ by measuring simultaneously the object endpoint ($x_n$-coordinates), that is to say $\Delta t_n=0$ in (\ref{r2E}), which gives the length contraction
\begin{equation}\label{stpn0}
\Delta x_n={l_0\over \gamma_n}.
\end{equation}

a) at each Step n, the length contraction is affected by the magnitude $v$ of the reference frame $\mathcal R'_n$ relative to $\mathcal R_n$.

b) moreover, the length contraction is also affected by the space-time expanding parameter. Indeed, since (\ref{stpn0}) is valid for all $n\geq 0$, then at the Step n+1 (for the same object at rest in the inertial reference frame  $\mathcal R'_{n+1}$ with the same proper length parallel to the $x_{n+1}$-axis given by $l_0=\Delta x'_{n+1}$), then the substitution of $\Delta t_{n+1}=0$ in ((\ref{r2E}), $T_{n+1}$) at the Step n+1, gives the length contraction
\begin{equation}\label{stpn1}
\Delta x_{n+1}={l_0\over \gamma_{n+1}},
 \end{equation}
then the ratio of (\ref{stpn1}) by (\ref{stpn0}) gives at successive Steps
\begin{equation}
\Delta x_{n+1}={\gamma_{n}\over \gamma_{n+1}}\Delta x_{n}
 \end{equation}
and since $\gamma_{n}(v)<\gamma_{n+1}(v)$, then  $\Delta x_{n+1}<\Delta x_{n}$.

\subsubsection{Limiting velocity for short and large scale of time $n$}
The transformations (\ref{E4g}) is defined for $1-{v^2\over c_n^2}>0$, that is to say for all $v<\di{c_n}$, where $c_n$ is a limiting velocity for moving bodies given by
\begin{equation}\label{vl}
c_{n}=\di{c\over \prod_{i=0}^n a_i},
\end{equation}
that depends on $c$ the fossil speed of light at the primordial universe, and the expanding parameter $\prod_{i=1}^n a_i$, which is an increasing and bounded product greater than one that makes the limiting velocity $c_{n}$ decreasing together with the space-time discrete expansion and it verifies:

\begin{prp}\label{light}
The sequence $(c_{n})_{n\geq0}$ given by (\ref{vl}) is constant for a short scale of time interval and variable for a large scale of time interval, with $c$ the constant speed of light recorded at the primordial space-time (Step 0).
\end{prp}

{\it Proof}: The properties of convergence of the expanding parameter $\prod_{i=0}^n a_i=e^{\sum_{i=0}^n \ln a_i}$ when n tends to the infinity means for all $\varepsilon>0$ there exists a big $N>0$ such that for $n\geq N$, $\prod_{i=0}^n a_i$ converge, then for all $n\geq N$, $\ln a_{n+1}$ tends to 0, that is to say $a_{n+1}$ tends to 1, then we have
\begin{equation}\label{Cv}
\prod_{i=0}^{n+1} a_i=(\prod_{i=0}^n a_i )\ a_{n+1}\approx\prod_{i=0}^n a_i,\qquad\forall n\geq N,
\end{equation}
thus, using (\ref{vl}), the equations (\ref{Cv}) gives
\begin{equation}\label{Cst}
c_{n+1}\approx c_{n}\qquad \forall n\geq N,
\end{equation}
that is to say the limiting velocity is constant for a short scale of time interval. Indeed the bigger $n$ is, the smaller the subdivision of time interval is (from primordial universe to present), then the smaller the considered scale of time after expansion is (see 2) in Remark\ref{rmak}).

However, the equation (\ref{Cst}) is not valid anymore for the large scale of time interval. Indeed, the smaller the number $n$ of steps is, the bigger the subdivision of time interval (from the primordial universe until present time) is, and in that case we have $a_{n+1}>1$ for all $n<N$, then
\begin{equation}\label{Cv0}
\prod_{i=0}^{n+1} a_i=(\prod_{i=0}^n a_i)\ a_{n+1}>\prod_{i=0}^n a_i,\qquad\forall n<N,
\end{equation}
and using (\ref{vl}) we have the following:
\begin{equation}
c_{n} > c_{n+1} \qquad \forall n<N,
\end{equation}
that is to say the limiting velocity is decreasing for a large scale of time interval.\pesp

\begin{rem}
1) Experimentally the measure of the speed of light with accuracy account since 1862 until today  (\cite{Gord},\cite{Even},\cite{Froo},\cite{LF},\cite{MICH1},\cite{MICH2},\cite{RD}), which indicates that the experimental measure of light was done within a short scale of time interval less than 200 years with respect to the age of the universe. These experimental measures gave us different values of the speed of light but all of them tend to the speed that account today for $2.99792458\times 10^8 m/s$. To measure the feasibility of a small variation of the speed of light for a large scale of time interval together with the space-time expansion, a measure of the speed of light must be compared during a large scale of time period, which is experimentally not accessible.

2) A limiting velocity that remains constant at a small scale of time and variable at a large scale of time interval means it is imperceptible at everyday measure, or every month or hundreds or thousands of years measure, but perceptible for a large scale of time measure.  However, if experimentally the variation of the limiting velocity is not perceptible even for a large scale of time period, then this approach in this case study remains a local approximation valid for a short scale of time. Nevertheless, a large scale observation in cosmology of phenomenon that presents a discrepancy between theory and observation might find its interpretation in this case study that takes into account consequences of a small variation of the limiting velocity for a large scale of time together with the space time expansion.
\end{rem}

\subsection{Physical interpretation of the space-time interval}

The space-time interval $\mathcal S_n^2$ given in the Definition \ref{Md} can be written using the differences between coordinates of a pair of events measured by an observer in $\mathcal R_n$ as follow:
\begin{equation}\label{pair}
\mathcal S^2_n=(c \Delta t)^2-(\prod_{i=0}^na_i)^2(\Delta x^2+ \Delta y^2+ \Delta z^2)=\prod_{i=0}^na_i[\Delta t(c_n-v)]\times \prod_{i=0}^na_i[\Delta t(c_n+v)],
\end{equation}
where the square magnitude of $v$ is given by $v^2=({\Delta x\over\Delta t})^2+ ({\Delta y\over\Delta t})^2+ ({\Delta z\over\Delta t})^2$.
Thus the space-time interval $\mathcal S_n^2$ for all $n\geq0$ is the areas $w\times l$ of a rectangle, as recorded by the observer in the reference frame $\mathcal R_n$ at the Step n, with width and length given by
$w=\prod_{i=0}^na_i[\Delta t(c_n-v)]$ and $l=\prod_{i=0}^na_i[\Delta t(c_n+v)]$ where $(\di c_n\pm v)$ is a difference or addition of magnitudes of velocities multiplied by an interval of time $\Delta t$ that provides a distance, and the multiplication of a distance by an expanding parameter is an expanding distance. The multiplication of $w\times l$ defines an area that measures how fast the event moves with respect to the limiting velocity $c_n$ by an observer in a reference frame $\mathcal R_n$ at the Step n for all $n\geq0$, and we have:

{\bf i)} If the area is equal to zero then the event moves at the speed of light $c_n$ as measured by an observer relative to a reference frame $\mathcal R_n$ at the Step n, $\forall n\geq0$,

 {\bf ii)}If the area is positive then the event moves at a speed less than the speed of light $c_n$ as measured by an observer relative to a reference frame $\mathcal R_n$ at the Step n, $\forall n\geq0$.

{\bf iii)} If the area is negative, then the motion of the event is impossible since the event move at a speed greater than the limiting velocity $c_n$ of moving bodies.

The invariance of the space-time interval means that the area $w\times l$ recorded by an observer in an inertial reference frame $\mathcal R_n$ is the same in any other inertial reference frame.

\section{Composition law}

\subsection{Composition law of velocity}

Consider the inertial reference frames $\mathcal R_n$ and $\mathcal R'_n$, where the frame $\mathcal R'_n$ moves with velocity $\vec{v}$ relative to the frame $\mathcal R_n$ parallel to their common $x$-axis. Consider a particle moving with constant velocity $\vec{u}$ as measured relative to the frame $\mathcal R_n$, and a constant velocity $\vec{u'}$ as measured relative to the frame $\mathcal R'_n$ at the Step n  for all $n\geq0$. The components of the velocity $\vec{u}$ and  the components  velocity $\vec{u'}$, in the differential limit, for all $n\geq0$ are given by
 \begin{equation}\label{CS}
 u_{x}={\Delta x\over\Delta t},\ u_{y}={\Delta y\over \Delta t},\ u_{z}={\Delta z\over \Delta t}\qquad\hbox{and}\qquad u'_{x_n}={\Delta x'\over \Delta t'},\ u'_{y}={\Delta y'\over \Delta t'},\ u'_{z}={\Delta z'\over \Delta t'}.
  \end{equation}

For simplicity and without losing generality consider only the case where the velocity $\vec{u}$ is parallel to the $x$ and $x'$-axis of frames $\mathcal R_n$ and $\mathcal R'_n$ for all $n\geq0$. The Lorentz transformations (\ref{E4g}) using the differences between coordinates for a pair of events as measured by observers in $\mathcal R_n$ as well as in $\mathcal R'_n$ at the Step n give for all $n\geq0$
\begin{equation}\label{E5}
T_n:\left\{
  \begin{array}{ll}
    \Delta x'= & \di{\Delta x-v\Delta t\over\sqrt{1-{v^2\over c_n^2}}} \\
    \Delta y'= & \Delta y \\
    \Delta z'= & \Delta z\\
    \Delta t'= & \di{\Delta t-{v\Delta x_n\over c_n^2}\over\sqrt{1-{v^2\over c_n^2}}}
  \end{array}
\right.\qquad\hbox{then}\qquad \left\{
  \begin{array}{ll}
    {\Delta x'\over \Delta t'}= & \di{\Delta x_-v\Delta t\over \Delta t-{v\over c_n^2}\Delta x} \\
    {\Delta y'\over \Delta t'}= & \di{\Delta y\over \Delta t-{v\Delta x\over c_n^2}}\Big(1-{v^2\over c_n^2}\Big)^{1\over2} \\
    {\Delta z'\over \Delta t'}= & \di{\Delta z\over \Delta t-{v\Delta x\over c_n^2}}\Big(1-{v^2\over c_n^2}\Big)^{1\over2} \\
  \end{array}
\right.
\end{equation}
and using (\ref{CS}), we have
\begin{equation}
\left\{
  \begin{array}{ll}
    u'_{x}={\Delta x'\over \Delta t'}= & \di{u_{x}-v\over 1-{v\over c_n^2}u_{x}} \\
    u'_{y}={\Delta y'\over \Delta t'}= & \di{u_{y}\over 1-{v\over c_n^2}u_{x}}\Big(1-{v^2\over c_n^2}\Big)^{1\over2} \\
    u'_{z}={\Delta z'\over \Delta t'}= & \di{u_{z}\over 1-{v\over c_n^2}u_{x}}\Big(1-{v^2\over c_n^2}\Big)^{1\over2}, \\
  \end{array}
\right.
\end{equation}
and since $\vec{u}$ is parallel to the $x$-axes, the law of composition of velocity for all $n\geq0$ is reduced to
\begin{equation}\label{u}
u'_{x'}= {u_{x}-v\over 1-{v\over c_n^2}u_{x}} \qquad\hbox{and inversely}\qquad u_{x}= {u'_{x'}+v\over 1+{v\over c_n^2}u'_{x'}}.
\end{equation}

\subsection{Invariance of the limiting velocity}

In the extreme case where an observer in the frame $\mathcal R'_n$ for all $n\geq0$ is emitting an event in the direction of $x$-axis with a velocity $u'_{x'}=c_{n}$, then using the inverse transformation of velocity (\ref{u}) for $u_x$, the velocity of this event in the frame $\mathcal R_n$ is given by
\begin{equation}
u_{x}= {c_{n}+v\over 1+{v\over c_n^2}c_{n}}={c_n+v\over 1+{v\over c_n}}={c_n+v\over {{c_n+v}\over c_n}}=c_{n}
\end{equation}
which suggests that the limiting velocity $c_{n}$ at the Step n for all $n\geq0$ is invariant (independent of the velocity of its source despite of its decreasing nature together with the space-time expansion), and represents the maximum speed of light that can be measured by an observer at each Step n for all $n\geq0$, and at present time (the Step n) the maximum speed of light is
\begin{equation}\label{C}
c_n=\di{c\over \prod_{i=0}^n a_i}=2.99792458 \times 10^8 \ m/s,
 \end{equation}
and this speed of light is constant if measured during a short scale of time interval with respect to the universe age, where c is the fossil speed of light as measured in the primordial universe (step 0) when it was liberated at the first time after the transformation of our universe from opaque to transparent.

\section{Momentum and relative mass}

The newtonian momentum is defined by $P(v)=mv=m{\Delta x\over\Delta t}$, where the average speed of the physical system corresponds to  $v={\Delta x\over\Delta t}$. To define the relativistic momentum consider two inertial reference frames $\mathcal R_n$ and $\mathcal R'_n$,  where the frame $\mathcal R'_n$ moves with velocity $\vec{v}$ relative to the frame $\mathcal R_n$ parallel to their common $x$-axis. Consider a particle at rest in the inertial reference frame $\mathcal R'_n$, then $\Delta x'=0$, which gives using $T^{-1}_n$ in (\ref{r2E}) the equality $\Delta t=\gamma_n\Delta t'$ for a time interval $\Delta t'\neq0$. Thus, the time interval of length $\Delta t $ is a linear function  of $\Delta  t'$ (the change of the time interval $\Delta t $ depends on the change of $\Delta t' $). Moreover, Using $T^{-1}_n$ in (\ref{r2E}) for $\Delta x'=0$ gives $\Delta x=\gamma_n v\Delta t'$, then \begin{equation}
m\Delta x=m\gamma_n v\Delta t'
 \end{equation}
 and the average change of the physical system displacement relative to an observer in the reference frame $\mathcal R_n$ depends on the rate of change of the time interval of length $\Delta  t'$. Then the change of the mass-displacement $m\Delta x$ over the minimal change of the time interval (which is the interval of time of length $\Delta  t'$ since $\Delta  t>\Delta  t'$) defines the relative momentum as recorded by the observer in the reference frame $\mathcal R_n$, then the particle's momentum has a magnitude defined by
\begin{equation}\label{mom}
P_n(v)=m{\Delta x\over\Delta t'}=m\gamma_n v={mv\over\sqrt{ 1-{v^2\over c_n^2}}},
\end{equation}
where $\Delta x$ is the covered distance by the particle as recorded by an observer relative to the reference frame $\mathcal R_n$, during the time $\Delta t'$ registered by an observer in the inertial reference frame $\mathcal R'_n$. The relative mass of the particle in motion, denoted by $m_n(v)$, as measured by an observer in the reference frame $\mathcal R_n$ is for all $n\ge0$ given by
\begin{equation}\label{rela}
m_n(v)={m\over\sqrt{1-{v^2\over c_n^2}}}=m\gamma_n(v)\qquad\hbox{and at rest}\qquad m_n(0)={m}.
\end{equation}

The formula (\ref{rela}) suggests that the relative mass $m_n$ is affected not only by the magnitude of the velocity $v$ of the particle but also by the space-time expanding parameter $\prod_{i=0}^n a_i$.
\subsection{Mass and momentum variation together with the space-time expansion}

The more the space-time expands, the more the momentum increases by virtue of the factor $\gamma_n$ (defined in (\ref{gn})) for a constant magnitude of the velocity $\vec{v}$. Indeed, using (\ref{mom}) and (\ref{rela}) we have the following
\begin{prop}
Consider a physical system moving in the space-time $\mathcal E$ with non null rest mass m and constant velocity with magnitude $v$ parallel to the x-axis of the inertial reference frame $\mathcal R_n$ at the Step n for all $n\geq0$, then for all $n\geq0$:

i) the relative mass (\ref{rela}) of the physical system is increasing together with the space-time discrete expansion.

ii) the relative momentum (\ref{mom}) of the physical system  is increasing together with the space-time discrete expansion.
\end{prop}

{\it Proof}: i) Indeed,  since the non null rest mass $m$ is invariant under space-time expansion then the equality (\ref{rela}) at the Step n and at the Step n+1 gives
\begin{equation}
m={m_{n+1}(v)\over \gamma_{n+1}(v)}={m_n(v)\over \gamma_n(v)},
\end{equation}
with $\gamma_n$ defined by (\ref{gn}) for all $n\geq0$, then for all $n\geq0$ such that $v<c_{n+1}$ we have
\begin{equation}
{m_{n+1}(v)\over m_n(v)}={\gamma_{n+1}(v)\over \gamma_{n}(v)},
\end{equation}
and since the sequence $(c_n)_{\geq0}$ is a decreasing then the sequence $(\gamma_n)_{\geq0}$ is increasing
which gives
\begin{equation}
{m_{n+1}(v)\over m_n(v)}={\gamma_{n+1}(v)\over \gamma_{n}(v)}>1,
\end{equation}
and then $\forall\ n\geq 0$, $m_{n+1}(v)>m_n(v)$.

ii) Since $\forall\ n\geq 0$, $m_{n+1}(v)>m_n(v)$, then a multiplication by the magnitude $v$ of the physical system velocity gives $\forall n\geq0$
\begin{equation}\label{Mc}
m_{n+1}(v)\cdot v>m_n(v)\cdot v,\qquad\hbox{then}\qquad P_{n+1}(v)>P_n(v)
\end{equation}
that is to say the relative momentum $P_n(v)$ is increasing together with the space-time discrete expansion for a physical system moving with a constant velocity of magnitude $v$ for all $n\geq0$.

\section{Energy}

\subsection{Kinetic energy and total energy}
The total energy $E=mc^2$ (Einstein \cite{EI3}) can be adjusted using the new limiting velocity and taking into account the space-time discrete  expansion:

\begin{prop}
Consider a physical system moving in the space-time $\mathcal E$ with non null rest mass m and constant velocity with magnitude $v$ parallel to the x-axis of the inertial reference frame $\mathcal R_n$ for all $n\geq0$, then for all $n\geq0$ the relativistic kinetic energy $K_{n}$ of the physical system is given by
\begin{equation}\label{Kinetic}
K_{n}=(m_n(v)-m)c_n^2,
\end{equation}
where the total energy $E_n$ and the rest-mass energy $E_{n}(0)$ of the physical system are given by
\begin{equation}\label{E6}
E_n(v)=m_n(v)\ c_n^2\qquad\hbox{and}\qquad E_{n}(0)=m\ c_n^2,
\end{equation}
with for all $n\geq0$, $m=m_n(0)$ is the rest mass, $m_n(v)$ is its relative mass defined by (\ref{rela}), and $c_n$ the speed of light defined by (\ref{vl}).
\end{prop}

{\it Proof:}
The proof is identical to the classical proof using (\ref{mom}) and (\ref{rela}) (see (\cite{Boratav}), \S 5.4).

\begin{rmak}
 The energy and momentum are related using formula (\ref{E6}) and (\ref{mom}) for all $n\geq0$ by
\begin{equation}\label{e1}
P_n\ c_n=E_n\ {v\over c_n}.
\end{equation}
\end{rmak}

\subsection{Energy variation together with the space-time expansion}

More effects of the discrete expansion on the physical laws can be figured out:
\begin{prop}\label{cinet}
Consider a physical system moving in the space-time $\mathcal E$ with non null rest mass m and a constant velocity with magnitude $v$ parallel to the x-axis of the inertial reference frame $\mathcal R_n$ for all $n\geq0$. If the magnitude $v\in]0,c_n[$ for all $n\geq0$  then the relativistic kinetic energy $K_n=m\ c_n^2(\gamma_n-1)$ of the physical system is increasing together with the space-time discrete expansion.
\end{prop}

{Proof:} The kinetic energy (\ref{Kinetic}) gives
\begin{equation}\label{KIN}
K_{n}=m\Big({1\over\sqrt{1-{v^2\over c_n^2}}}-1\Big)c_n^2=m\ v^2f(u_n)
\end{equation}
where for $x\in]0,1[$
\begin{equation}\label{deri1}
f(x)=\Big({1\over \sqrt{1-x^2}}-1\Big){1\over x^2} \qquad \hbox{and} \qquad u_n={v\over c_n}.
\end{equation}

\begin{figure}[!h]
\centering
\includegraphics[width=6cm]{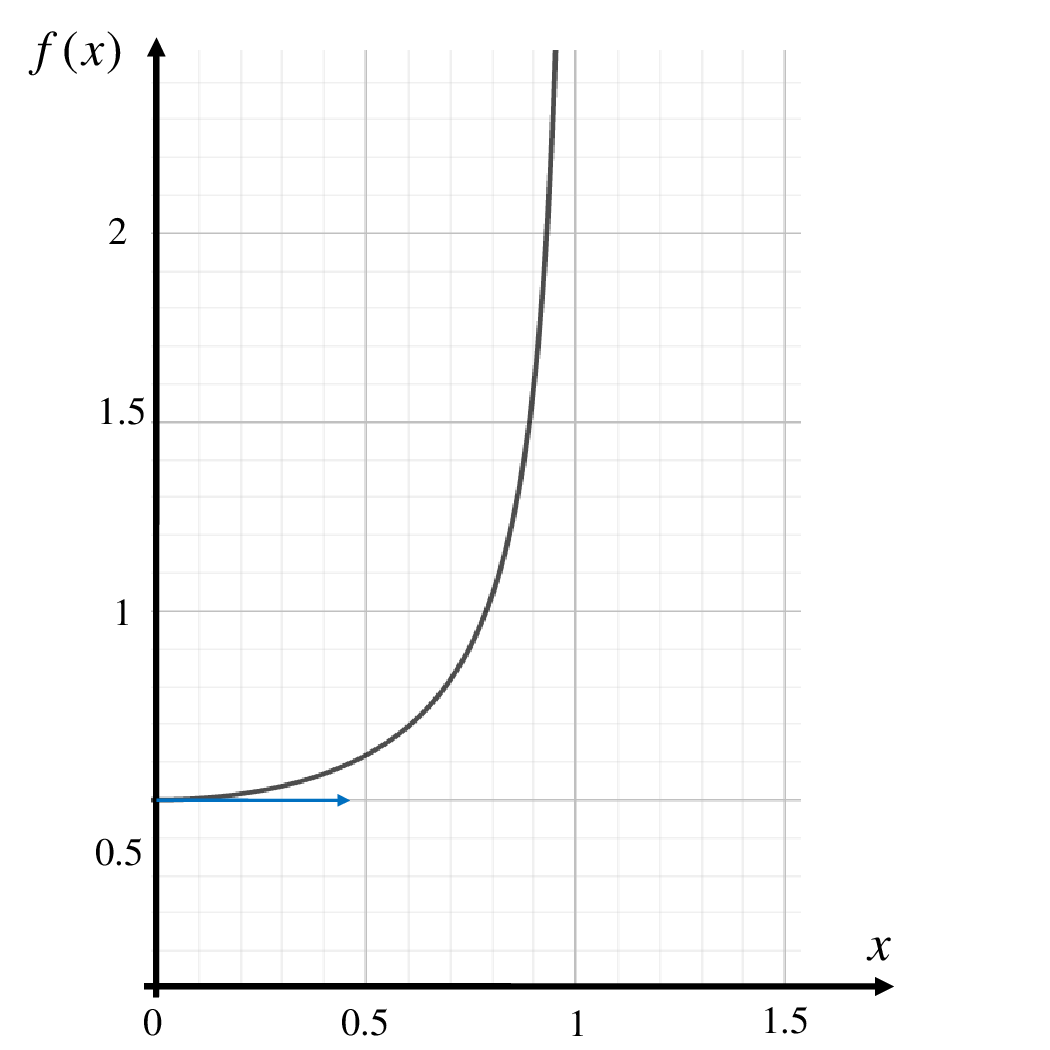}
\caption{\footnotesize The graph of the function $f(x)=({1\over \sqrt{1-x^2}}-1){1\over x^2}$ for $x\in ]0,1[$. The function $f$ is increasing for $x\in]0,1[$. Its graph has a vertical asymptote $x=1$ and horizontal tangent at $(0,{1\over2})$. The function $f(x)$ is not defined at $x=0$ but $\lim_{x\rightarrow 0}f(x)=0.5$.}\label{Fig.9}
\end{figure}

Since the sequence $(c_n)_{n\geq0}$ is decreasing, then sequence $(u_n)_{n\geq0}$ is increasing and the first derivative of the function $f$ for all $0<x<1$ is given by
\begin{equation}
f'(x)={3x^2-2+(2-2x^2)\sqrt{1-x^2}\over x^3\sqrt{1-x^2}^3},
\end{equation}
and $f'(x)>0$ for all $0<x<1$. Thus, $f$ is an increasing function in $]0,1[$ (see Fig.\ref{Fig.9}).
Moreover if for all $n\geq0$ the magnitude $v\in]0,c_n[$, then $0<{v\over c_n}<1$, that gives $0<u_n<1$, and since the sequence $(c_n)_{n\geq0}$ is decreasing then the sequence $(u_n)_{n\geq0}$ is increasing together with the space-time discrete expansion (ie. for all $n\geq0$, $u_{n+1}>u_n$). Therefore the increase of the function $f$ in $]0,1[$ gives
\begin{equation}
K_{n+1}-K_n=m\ v^2\Big(f(u_{n+1})-f(u_n)\Big)>0,\quad\hbox{then}\quad\forall n\geq0,\quad K_{n+1}>K_n.
\end{equation}

\begin{prop}\label{restE}
The rest energy $E_{n}(0)=m\ c_n^2$ of the physical system introduced in (\ref{E6}) is decreasing together with the space-time discrete expansion.
\end{prop}

{\it Proof}: The rest energy at the Step n can be evaluated compared to the rest energy at the Step 0 (the primordial space-time). Indeed the rest energy at the Step n
\begin{equation}\label{En}
E_{n}(0)=m\ c_n^2={mc^2\over \prod_{i=0}^n a^2_i}={E_0(0)\over \prod_{i=0}^n a^2_i},
\end{equation}
with $E_0(0)=mc^2$ the rest energy at the Step 0, which gives ${E_{n}(0)\over E_{n+1}(0)}=a^2_{n+1}$, and
since $a_{n+1}>1$ (Definition \ref{para}), then for all $n\geq0$, ${E_{n}(0)>E_{n+1}(0)}$.

\begin{rmak}
The equality (\ref{En}) allows to evaluate the lost rest energy together with the space-time discrete expansion between the Step 0 and the Step n (present time), indeed
\begin{equation}\label{En0}
E_n(0)- E_0(0)=E_0(0)\Big({1\over \prod_{i=0}^n a_i^2}-1\Big)
\end{equation}
\end{rmak}

\begin{prop}
Consider a physical system moving in the space-time $\mathcal E$ with non null rest mass m and constant velocity with magnitude $v$ parallel to the x-axis of the inertial reference frame $\mathcal R_n$ for all $n\geq0$, then for all $n\geq0$,

i) If the magnitude $v\in [0,\sqrt{2\over3}c_n[$, then the total energy $E_n=m_n(v)c_n^2$ is decreasing together with the space-time discrete expansion.

ii)  If  the magnitude $v\in [\sqrt{2\over3}c_n,c_n[$ then the total energy $E_n=m_n(v)c_n^2$ is increasing together with the space-time discrete expansion.

\end{prop}

{\it Proof}: The total energy (\ref{E6}) of the physical system can be expressed as
\begin{equation}\label{Total}
E_n=m{1\over\sqrt{1-{v^2\over c_n^2}}}c_n^2=mv^2\times f(u_n)
\end{equation}
where for $x\in]0,1[$
\begin{equation}\label{deri2}
f(x)={1\over x\sqrt{1-x}} \qquad \hbox{and} \qquad u_n={v^2\over c_n^2}.
\end{equation}

\begin{figure}[h]
\centering
\includegraphics[width=6cm]{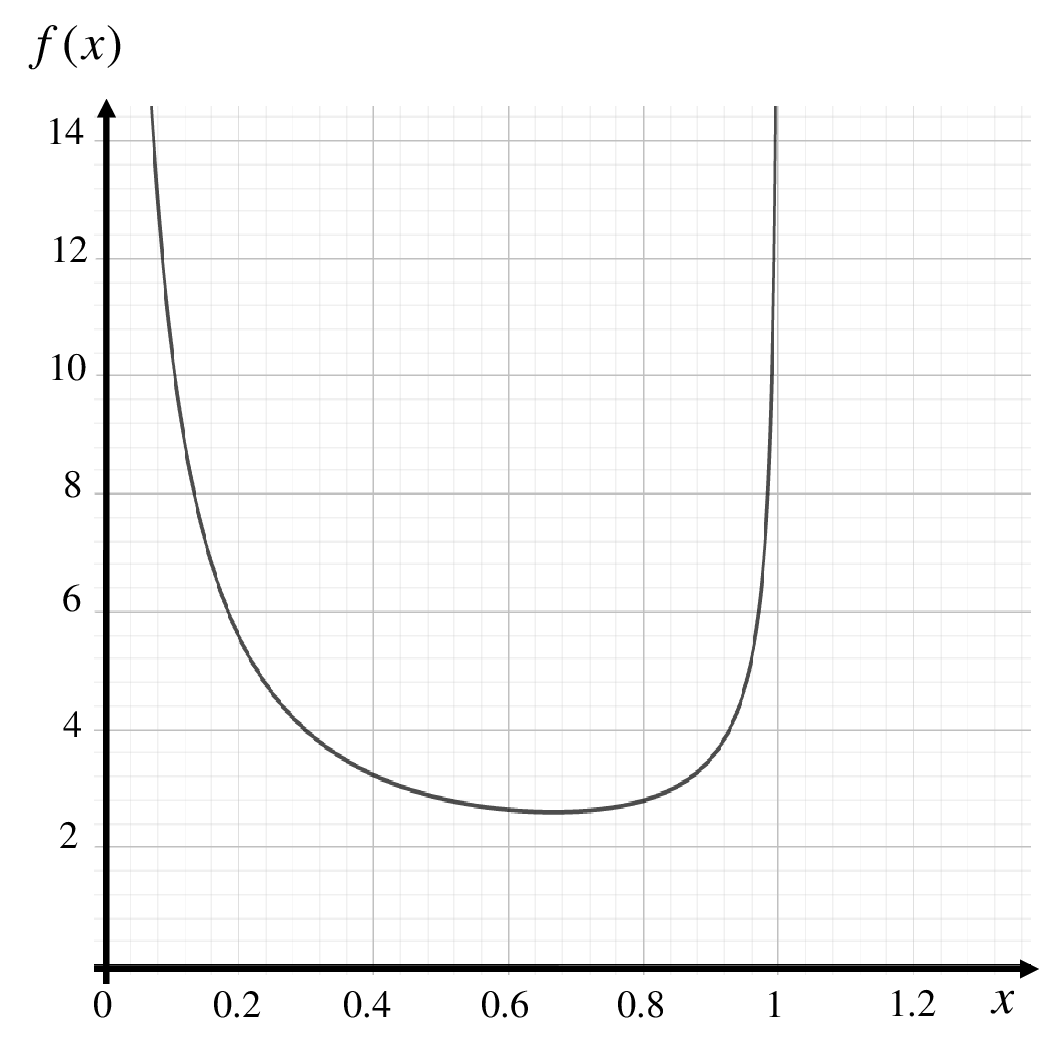}
\caption{\footnotesize The graph of the function $f(x)={1\over x\sqrt{1-x}}$ for $x\in ]0,1[$ has two vertical asymptotes at $x=0$ and $x=1$ and  horizontal tangent at $({2\over3},f({2\over3}))$ . The total energy $E_n=m\ v^2f(u_n)$ for $u_n={v^2\over c_n^2}$ and the function $f$ share the same monotony for a constant velocity of magnitude $0<v<c_n$. }\label{Fig.10}
\end{figure}

Since $(c_n)_{n\geq0}$ is decreasing sequence, then the sequence $(u_n)_{n\geq0}$ is an increasing sequence. The monotonic of (\ref{Total}) together with the space-time discrete expansion is determined by the monotonic of the function $f$. Indeed, the derivative of the function $f$ for all $0<x<1$ gives
\begin{equation}\label{deriv}
f'(x)={3x-2\over2x^2\sqrt{1-x}^3}
\end{equation}
where $f'(x)<0$ for $0<x<{2\over3}$ and $f'(x)>0$ for ${2\over3}<x<1$,
meanwhile $f'({2\over3})=0$, that is to say the function $f$ is decreasing for $0<x<{2\over3}$ and increasing for ${2\over3}<x<1$ and its graph has horizontal tangent at the point $({2\over3}, f({2\over3}))$ (see Fig.\ref{Fig.10}). Then

i)  If for all $n\geq0$ the magnitude $v\in]0,\sqrt{2\over3}c_n[$, then for all $n\geq 0$,  $0<{v\over c_n}<\sqrt{2\over3}$, and using (\ref{deri2}) it gives  $0<u_n<{2\over3}$. Hence, since the sequence $(c_n)_{n\geq0}$ is decreasing then the sequence $(u_n)_{n\geq0}$ is increasing together with the space-time discrete expansion (ie. for all $n\geq0$, $u_{n+1}>u_n$), moreover since the function $f$ is decreasing in $]0,{2\over3}[$ then
\begin{equation}
E_{n+1}(v)-E_n(v)=mv^2 \Big(f(u_{n+1})-f(u_n)\Big)<0
\end{equation}
and then for all $n\geq0$, it gives $E_{n+1}(v)<E_n(v)$. For $v=0$, the rest energy $E_n(0)$ is decreasing together with the space-time discrete expansion (see Proposition \ref{restE}), which concludes the proof of i)

ii)  If for all $n\geq0$ the magnitude $v\in [\sqrt{2\over3}c_n,c_n[$,  then for all $n\geq 0$, $\sqrt{2\over3}\leq{v\over c_n}<1$, and using (\ref{deri2}) it gives ${2\over3}\leq u_n<1$. Thus the sequence $(u_n)_{n\geq0}\in\ [{2\over3},1[$, and verifies for all $n\geq0$, $u_{n+1}>u_n$, moreover since the function $f$ is increasing in $]{2\over3},1[$ then
\begin{equation}
E_{n+1}(v)-E_n(v)=mv^2 \Big(f(u_{n+1})-f(u_n)\Big)>0
\end{equation}
 and then for all $n\geq0$, it gives $E_{n+1}(v)>E_n(v)$ which completes the proof of ii).

\begin{rem}\label{R9}
1) At present time (Step n) the value $\sqrt{2\over3}c_n$ is a critical speed and is given  by the numerical value
\begin{equation}\label{cri}
\sqrt{2\over3}c_n=\sqrt{2\over3}\times 2.99792458 \times 10^8 \ m/s=244779516.9\ m/s.
\end{equation}

At this critical speed, the total energy (\ref{E6}) changes its monotonic from a decreasing total energy to an increasing total energy together with the space-time discrete expansion, which allows based on this boundary to classify the physical systems into two categories at present time:

a) The category of physical systems with constant speed of magnitude
\begin{equation}
v\ \in\ \Big[0\ ,\ 2.447795169\times10^8\Big[
\end{equation}
where their total energy decreases together with the space-time expansion.

b) The category of physical systems with constant high speed of magnitude
\begin{equation}\label{Hspeed}
v\ \in\ \Big[2.447795169\times 10^8\ ,\ 2.99792458\times  10^8 \Big[
\end{equation}
 where their total energy increases together with the space-time expansion.

2) Moreover the critical speed $\sqrt{2\over3}c_n$ is decreasing together with the space-time expansion.
\end{rem}

\section{Time interval together with the space-time expansion}

Consider two inertial reference frames $\mathcal R'_n$ and $\mathcal R_n$ in the space-time $\mathcal E$, where the inertial reference frame $\mathcal R'_n$ has a constant velocity with magnitude $v$, relative to the frame $\mathcal R_n$ in the positive $x$-direction such that the corresponding axes of $\mathcal R_n$ and $\mathcal R'_n$ remain parallel throughout the motion, and where the origins of the frames having coincided at $t_n=t'_n=0$ for all $n\geq 0$ (the last equality is possible for all $n\geq 0$ if the events are repeated n times under the same condition for all $n\geq 0$).

We refer the proper time interval $\Delta\tau_n$ for all $n\geq0$ as the temporal separation between two events, as measured by an observer in the inertial reference frame $\mathcal R'_n$, that occur at the same location.

\subsection{Low kinetic energy and proper time dilation}

Consider two events that involve a physical system with non null kinetic energy and that occur at the same spatial point relative to the inertial reference frame $\mathcal R'_n$ for all $n\geq 0$.  Then we have:
\begin{prop}\label{CSTPT}
If the physical system's speed satisfies $v\ll c_n$, then the kinetic energy of the physical system and its proper time are  invariant together with the space-time discrete expansion.
\end{prop}

{\it Proof}: Using the function (\ref{deri1}),  the kinetic energy (\ref{Kinetic}) is given by $K_{n}=m\ v^2f(u_n)$,
where for all $n\geq0$, $u_n={v\over c_n}$. Since the function (\ref{deri1}) admits an horizontal tangent at $(0,{1\over2})$ (Fig.\ref{Fig.9}), then for $v\ll c_n$, there exists a right side neighborhood $V^+(0)$ such that for all $u_n<u_{n+1}$ in $V^+(0)$ we have $K_{n+1}-K_n=m\ v^2(f(u_{n+1})-f(u_n)\Big )\approx0$, that is to say $K_{n+1}=K_n={1\over2}m\ v^2$ which is invariant together with the space-time discrete expansion (not affected by the space expansion).

\begin{figure}[!ht]
\begin{minipage}[t]{6cm}
\centering
\includegraphics[width=6cm]{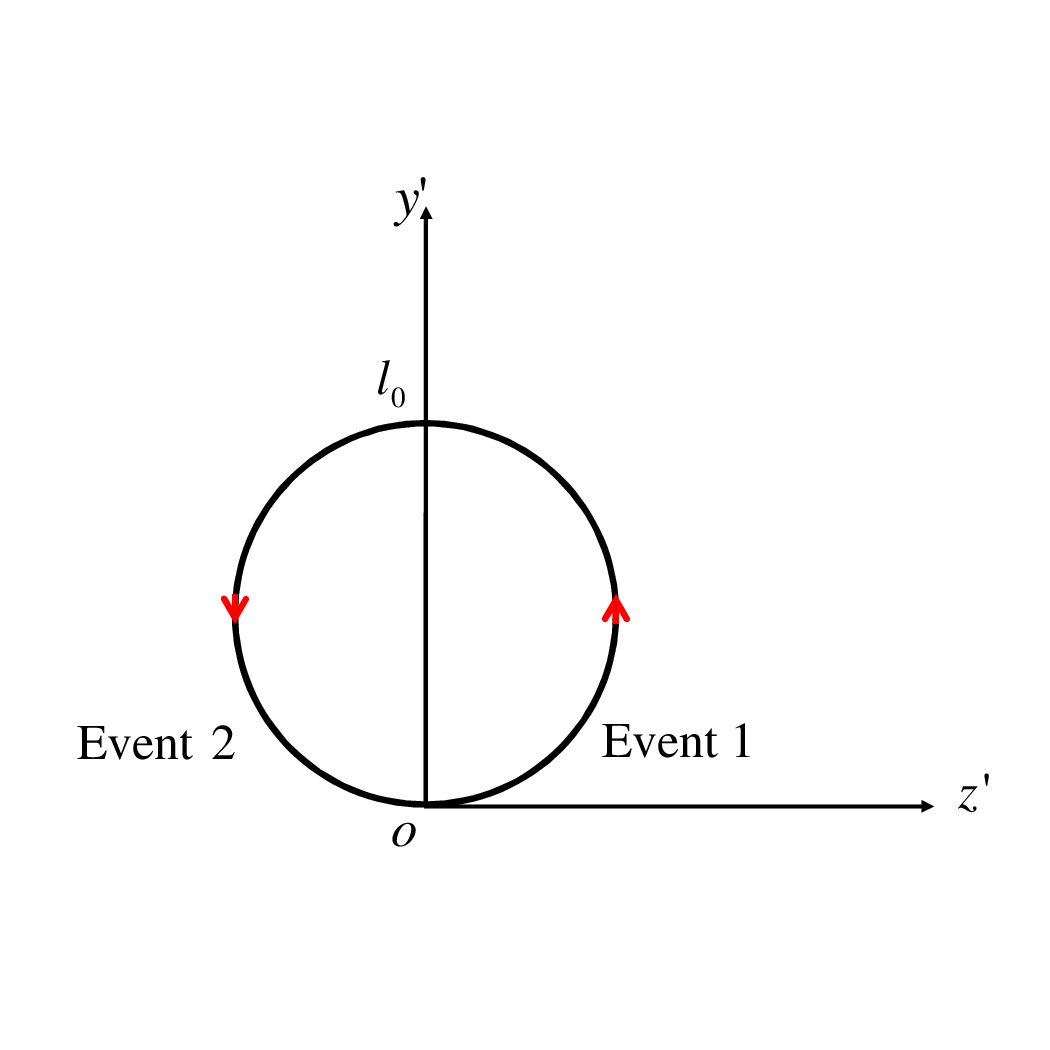}
\caption{\footnotesize A circle $\cal C$ of center $(0,{l_0\over2},0)$ and equation  $(y-{l_0\over2})^2+z^2={l_0^2\over4}$ in the $yz$-plan. The intersection ${\cal C}\cap \{y=0\}=(0,0,0)$.}\label{Fig.001}
\end{minipage}
\hspace*{\fill}
\begin{minipage}[t]{6cm}
\centering
\includegraphics[width=6cm]{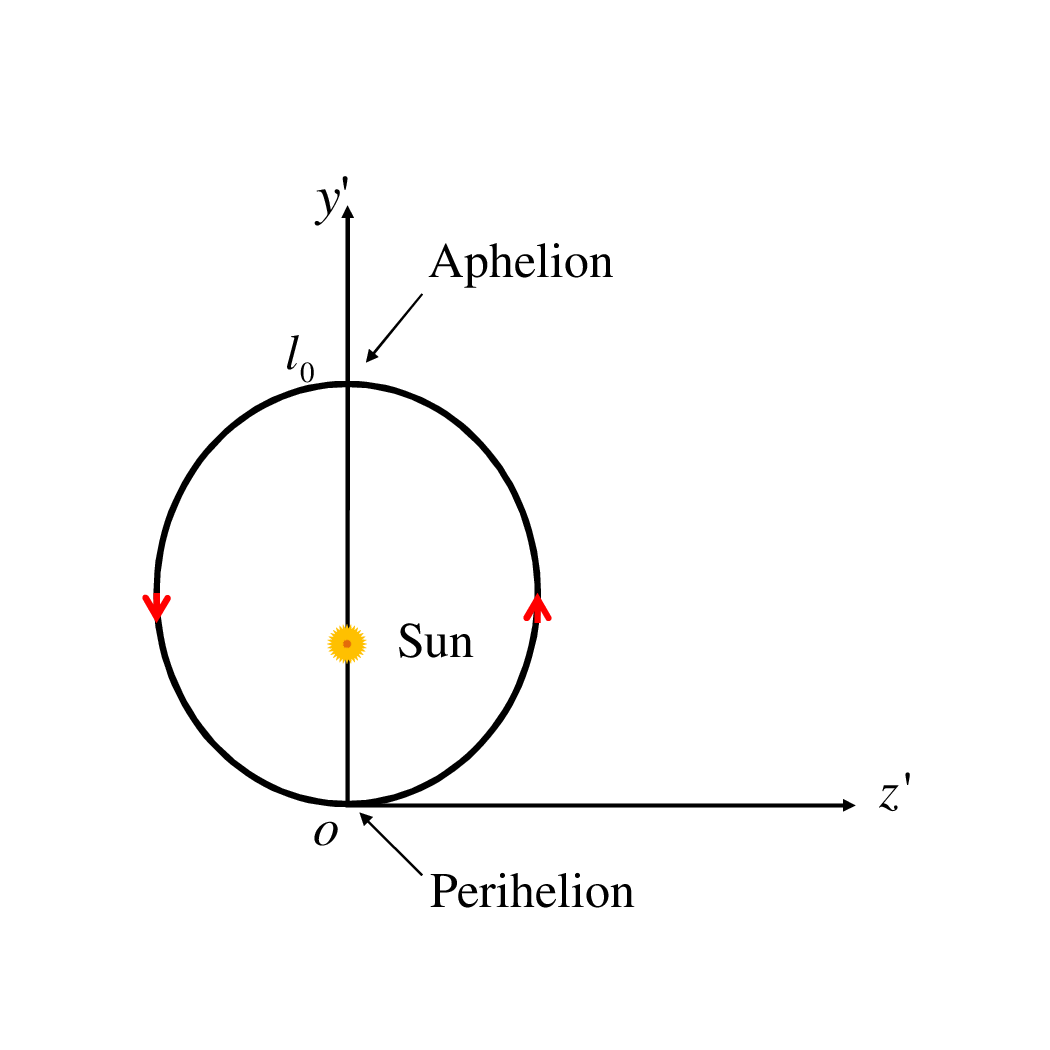}
\caption{\footnotesize Counterclockwise Earth orbit, where the Sun is at rest in the inertial reference frame $\mathcal R'_n$. The perihelion is located at the origin of the frame $\mathcal R'_n$.}\label{Fig.0002}
\end{minipage}
\end{figure}

To verify the invariance of the proper time for all $n\geq0$, consider two events that occur at the same location involving a physical system following an uniform vertical circular motion $\cal C$ of fix diameter $l_0$ in the $y'z'$-plan of the inertial reference frame $\mathcal R'_n$, with a speed of magnitude $0<v\ll c_n$ for all $n\geq0$, and where the origin of the frame $\mathcal R'_n$ coincides with the intersection point of $\cal C$ and the $x'z'$-plan (Fig.\ref{Fig.001}).
 The observer records the coordinates of two events: the departure of the physical system from the origin, and the return to it, and reports  $\Delta x'_n=\Delta y'_n=\Delta z'_n=0$,  $\Delta t'_n=\Delta\tau_n$ (non null proper time). Since the kinetic energy for the physical system is invariant together with the space-time discrete expansion, then one revolution of the physical system around $y={l_0\over2}$ at the Step n and the Step n+1 gives $l_0\pi=v\Delta\tau_n=v\Delta\tau_{n+1}$, which gives that for all $n\geq0$,  $\Delta\tau_n=\Delta\tau_{n+1}$ and completes the proof.
\begin{prop}
Consider two inertial reference frames $\mathcal R'_n$ and $\mathcal R_n$ in the space-time $\mathcal E$, where the inertial reference frame $\mathcal R'_n$ has a constant velocity with magnitude $v$, relative to the frame $\mathcal R_n$ in the positive $x$-direction for all $n\geq0$, and where the origins of the frames having coincided at $t_n=t'_n=0$ for all $n\geq 0$. If an observer, in the frame $\mathcal R'_n$, reports a non null proper time $\Delta\tau_n$ for two events that occur at the same location involving a physical system with a speed satisfying $v\ll c_n$ for all $n\geq0$, then the relative time of the events as reported by an observer in the inertial reference frames $\mathcal R_n$  and $\mathcal R_{n+1}$ verifies for all $n\geq 0$
\begin{equation}\label{Ttau}
\Delta t_{n+1} ={\gamma_{n+1}(v)\over\gamma_n(v)}\Delta t_{n}
\end{equation}
with $\gamma_n$ defined by (\ref{gn}).
\end{prop}

{\it Proof:} if an observer in the rest frame $\mathcal R'_n$ reports a non null proper time $\Delta t'_n=\Delta \tau_n\neq0$ for the events that occur at the same location for all $n\geq0$, then the observer reports $\Delta x'_n=\Delta y'_n=\Delta z'_n=0$ and $\Delta t'_n=\Delta\tau_n$ at the Step n, and $\Delta x'_{n+1}=\Delta y'_{n+1}=\Delta z'_{n+1}=0$ and $\Delta t'_{n+1}=\Delta\tau_{n+1}$ at the Step n+1 (if the experiment is repeated at the Step n+1 under the same conditions). The substitution of the space-time coordinates at the Step n and the Step n+1 in the Lorentz transformation $T_n^{-1}$ (equality (\ref{r2E})) gives for all $n\geq0$
\begin{equation}\label{TP}
\Delta t_n=\gamma_n(v)\Delta\tau_n,\qquad\hbox{and}\qquad \Delta t_{n+1}=\gamma_{n+1}(v)\Delta\tau_{n+1},
\end{equation}
and since the events involving a physical system with $v\ll c_n$ for all $n\geq0$, then  $\Delta\tau_n=\Delta\tau_{n+1}$ (Proposition \ref{CSTPT}), and a substitution in (\ref{TP}) gives
\begin{equation}
\left\{
  \begin{array}{ll}
    \Delta\tau_n= & {\Delta t_n\over \gamma_n(v)} \\
    \Delta\tau_{n}= & {\Delta t_{n+1}\over \gamma_{n+1}(v)}
  \end{array}
\right.
\qquad \Rightarrow\qquad 1={\Delta t_{n+1}\over \Delta t_{n}}{\gamma_n(v)\over\gamma_{n+1}(v)},
\end{equation}
which leads to (\ref{Ttau}) and concludes the proof.

\begin{rmak}
 Since $\gamma_{n+1}(v)>\gamma_n(v)$ then the previous result (\ref{Ttau}) gives $\Delta t_{n+1}>\Delta t_{n}$ for all $n\geq0$ which is a time dilation together with the space-time expansion. Time will be perceived running slower for everything moving (with low kinetic energy) relative to others together with the space-time discrete expansion.
\end{rmak}

\subsection{High kinetic energy and proper time dilation}
Consider two events that occur at the same location relative to the inertial reference frame $\mathcal R'_n$ for all $n\geq0$ that involve a physical system with high kinetic energy (a particle moving with high speed $c_n$), where the two events consist of a pulse of light sent from a source at the origin of the inertial reference frame $\mathcal R'_n$  in the positive $y'$-direction (Event 1), to a mirror perpendicular to the $y'$-axis and located at a fix distance $l_0$ for all $n\geq0$. The signal is reflected vertically downward (event 2) and detected back in the source at the origin of the frame $\mathcal R'_n$ for all $n\geq0$. Then an observer in the inertial reference frame $\mathcal R'_n$ for all $n\geq0$ reports that the two events are located in the same spatial coordinates. Consider a second inertial reference frames $\mathcal R_n$ in the space-time $\mathcal E$, where the inertial reference frame $\mathcal R'_n$ has a constant velocity with magnitude $v$, relative to the frame $\mathcal R_n$ in the positive $x$-direction for all $n\geq0$, and where the origins of the frames $\mathcal R_n$ and $\mathcal R'_n$ having coincided at $t_n=t'_n=0$ for all $n\geq 0$ (repeated motion at each Step). Then  we have the following:

\begin{prop}
The proper time of the above two events (involving a physical system with high kinetic energy) that occur at the same location as reported by observers in the inertial reference frames $\mathcal R'_n$ and $\mathcal R'_{n+1}$ (if repeated) verifies:
\begin{equation}\label{Tt2}
   \Delta\tau_{n+1}={c_{n}\over c_{n+1}}\Delta \tau_{n},
\end{equation}
and the relative time of the events as reported by observers in the inertial reference frames $\mathcal R_n$ and $\mathcal R_{n+1}$ (if repeated) verifies
\begin{equation}\label{Tt1}
\Delta t_{n+1}= {c_{n}\over c_{n+1}}{\gamma_{n+1}(v)\over \gamma_{n}(v)}\Delta t_{n}.
\end{equation}
\end{prop}

{\it Proof}: 1) If the two events occur at the same location in the reference frame $\mathcal R'_n$ for all $n\geq0$, then the observer in the inertial reference frame $\mathcal R'_n$ reports $\Delta x'_n=\Delta y'_n=\Delta z'_n=0$ and $\Delta t'_n=\Delta\tau_n\neq0$ for all $n\geq0$. The substitution of $\Delta x'_n=0$ in (\ref{r2E}) using $T_n^{-1}$ (respectively the substitution of $\Delta x'_{n+1}=0$ in (\ref{r2E}) using $T_{n+1}^{-1}$) gives for an observer in the reference frame $\mathcal R_n$ (respectively in the inertial frame $\mathcal R_{n+1}$)
\begin{equation}\label{TP1}
\Delta t_n=\gamma_n(v)\Delta \tau_n\qquad\hbox{and}\qquad \Delta t_{n+1}=\gamma_{n+1}(v)\Delta \tau_{n+1}.
\end{equation}

 The emission and reflection of a pulse of light for a fix distance $l_0$ at the Step n and at the Step n+1 gives
\begin{equation}\label{l0}
2l_0=c_n\Delta\tau_n=c_{n+1}\Delta \tau_{n+1},
\end{equation}
which gives
\begin{equation}
\Delta\tau_{n+1}={c_{n}\over c_{n+1}}\Delta \tau_{n},
\end{equation}

The use of (\ref{TP1}) at the Step n and Step n+1 (under the same conditions) with (\ref{l0}) gives
\begin{equation}
2l_0=c_n{\Delta t_n\over\gamma_n(v)}= c_{n+1}{\Delta t_{n+1}\over\gamma_{n+1}(v)},
\end{equation}
then
\begin{equation}\label{Timedilation}
\Delta t_{n+1}= {c_{n}\over c_{n+1}}{\gamma_{n+1}(v)\over \gamma_{n}(v)}\Delta t_{n}.
\end{equation}

\begin{rem}
i) Since ${c_{n+1}\over c_n}<1$, then equality (\ref{Tt2}) indicates that the proper time of the physical system with high kinetic energy experiences proper time dilation together with the space-time discrete expansion and we have for all $n\geq0$
\begin{equation}\label{Dilation0}
\Delta \tau_{n}<\Delta \tau_{n+1},
\end{equation}
and since ${\gamma_{n}(v)\over \gamma_{n+1}(v)}<1$, then equality (\ref{Tt1}) indicates that the time dilation is more pronounced together with the space-time expansion and we have
\begin{equation}\label{Dilation}
\Delta t_n<\Delta t_{n+1}.
\end{equation}

ii) The variation of the proper time (\ref{Dilation0}) does not concern only events of physical system with the speed of light $c_n$. Indeed, the proper time variation concerns also any physical system with high speed close to the speed of light (near the asymptote) since the decrease of the limiting velocity (\ref{vl}) together with the space-time expansion will decrease the closer physical system's speed (see Example 1 in the Application below), which induces a variation of their proper time.
\end{rem}

\subsection{New insight for time}\label{New}

\begin{table}[!h]
\caption{Relative time and proper time for physical system with high kinetic energy and low kinetic energy using (\ref{TP}), (\ref{Ttau}), (\ref{Tt2}), (\ref{Tt1}), (\ref{Dilation0}), and (\ref{Dilation}). }
\label{T11}
\center\begin{tabular}{|l|l|l|l|}
\hline
 $\forall n\geq0$ &
$\left.
  \begin{array}{ll}
   {\bf \Delta t_n=f(\Delta\tau_n)}&\\
   ({\bf 0\leq v< c_n})&
  \end{array}
\right.$ & $\bf \Delta t_{n+1}=f(\Delta t_{n})$ & {\bf Time} \\
\hline
{\bf Step n} &$\Delta t_n=\gamma_n(v)\Delta\tau_n$ &  NA &$\Delta\tau_n\leq\Delta t_{n}$\\
\hline
$\left.
  \begin{array}{ll}
   \hbox{\bf Low K. Eng.}  &\\
   n\rightarrow n+1 &
  \end{array}
\right.$
 &
\small$\Delta t_{n}=\gamma_n(v)\Delta\tau_n$
 & $\left.
  \begin{array}{ll}
\small\Delta t_{n+1}={\gamma_{n+1}\over\gamma_n}\Delta t_n&\\
(\small\hbox{for}\ \Delta\tau_n= \Delta\tau_{n+1}) &
  \end{array}
\right.$
& $\left.
  \begin{array}{ll}
\small\Delta t_{n}\leq\Delta t_{n+1}&\\
\small\Delta \tau_{n}=\Delta\tau_{n+1}&
  \end{array}
\right.$\\
\hline
 $\left.
  \begin{array}{ll}
   \hbox{\bf High K. Eng.}&\\
   n\rightarrow n+1 &
  \end{array}
\right.$
   &
   \small$\Delta t_n=\gamma_n(v)\Delta\tau_n$
  & \small $\left.
  \begin{array}{lll}
 \small  \Delta \tau_{n+1}={c_{n+1}\over c_n}\Delta\tau_n   &\\
 \small \Delta t_{n+1}={c_{n+1}\over c_n}{\gamma_{n+1}\over\gamma_{n}}\Delta t_{n} &\\
 (\small\hbox{for}\ \Delta \tau_{n}\neq\Delta\tau_{n+1}) &
  \end{array}
\right.$ & $\left.
  \begin{array}{ll}
   \small \Delta t_{n}<\Delta t_{n+1} &\\
\small \Delta \tau_{n}<\Delta\tau_{n+1}&
  \end{array}
\right.$ \\
\hline
\end{tabular}
\vspace*{2pt}
\end{table}
 Based on the Lorentz transformations (\ref{E4g}) the measure of time is relative to the observer motion and the space expansion, where the rate of time runs differently depending on the physical system kinetic energy (Table \ref{T11}). Indeed two cases are figured out:

i) for physical system with low kinetic energy the proper time is invariant with the space-time discrete expansion, and the measure of relative time is affected  by motion and the space-time expansion (time dilation $\Delta t_{n+1}={\gamma_{n+1}\over\gamma_{n}}\Delta t_{n} $ in Table \ref{T11}).

ii) for physical system with high kinetic energy the proper time is affected  by the space-time discrete expansion (proper time dilation $\Delta \tau_{n+1}={c_{n+1}\over c_n}\Delta \tau_n$ in Table \ref{T11}), meanwhile the measure of the relative time is affected  by motion and space-time expansion. The time dilation is more pronounced by the virtu of space-time expansion ($\Delta t_{n+1}={c_{n+1}\over c_n}{\gamma_{n+1}\over\gamma_{n}}\Delta t_{n}$ ).

The proper time for physical systems with high kinetic energy appear to be more sensitive to the effect of space-time discrete expansion, meanwhile time dilation is imperceptible for a physical system with low level of kinetic energy. This particular sensitivity might reveal the nature of the course of time in an expanding space-time.

\subsubsection{Interpretation of time characteristics and description}\label{Time}
Based on this framework, interpretation of time can be elaborates as follow:

{\ni\bf Description:} time is a fundamental positive scalar (of one dimension) that indicates order of occurrence of all gradual changes of physical properties and  positions of matter and energy together with the three-dimensional space expansion (transformation).

{\ni\bf Duration, unit:} the scalar variable $t$ induces the duration (a quantity that measures the distance in time between two event's states) through the ratio of the time interval of the event's states (a temporal separation between start and end) to the time interval of an arbitrary number of occurrences of a repeating standard event (as a standard unit). The standard unit of measure depends on the oscillation (a repeating event) of the chosen motion of reference, and it can be affected by an external acceleration (if the motion of reference concerns a physical system with high energy), which makes the duration's measure relative to external acceleration.

{\ni\bf Implicit Definition:} time is defined implicitly by the continuum space-time interval (\ref{M1}) where it appears to be entangled to the space and its expansion, and it has existed since the existence of all gradual changes of physical properties and positions of matter and energy together with the three-dimensional space expansion.

{\ni\bf Course of time:} the course of time is represented by an oriented linear half axis that registers chronology of past events' states and indicates the future. Our movement thought time is ineluctable. We do not move through time but time records our gradual changes of physical properties and positions since our existence until our disappearance. We use a periodic standard motion of reference to measure duration of all local gradual changes of physical properties and position of matter and energy together with the three-dimensional space expansion for an arbitrary choice of time origin, with relative accuracy according to the observer motion and the space expansion (Table \ref{T11}).

{\ni\bf Irreversibility:} the measure of the course of time is directly attached to the unit of the standard motion of reference in the expanding space, which makes it unstoppable and irreversible since one cannot reverse the gradual changes of physical properties and positions of matter and energy together with the three-dimensional space expansion (transformation).

{\ni\bf Global invariance:} time does not generate events. An isolated physical system undergoes gradual changes of physical properties and positions function of time, and time runs independently of the isolated physical system's gradual changes.
For a reference frame at rest in the expanding space-time, the proper time of a physical system with low kinetic energy is invariant together with the space-time expansion (Proposition \ref{CSTPT}) meanwhile the proper time of a physical system with high kinetic energy experiences a continuous time dilation together with the space-time expansion (\ref{Tt2}).

{\ni\bf Second:} the local measurement of duration is relative to the observer motion in the expanding space and carried out using local clock that requires the definition of a unit of time: the second, and based on this framework the second must be an invariant unit of measure together with the space-time expansion.

\begin{rem}
1) In the 13th general conference on weights and measures (1967), it has been adopted a standard definition of second based on the cesium clock : one second is defined as the time taken by 9 192 631 770 oscillations of the light emitted by a cesium 133 atom (by a specific wave length) (\cite{WLK}). Based on this framework the proper time of physical system with high kinetic energy is affected by the space-time expansion (\ref{Tt2}). Indeed, the decrease of the sequence $(c_n)_{n\geq0}$ in  equality (\ref{Tt2}) induces the proper time dilation (\ref{Dilation}) of the physical system together with the space-time expansion,  therefore the 9 192 631 770 oscillations of the light emitted by a cesium 133 atom will take more time together with the space-time expansion (the flow of one second experiences a dilation). Meanwhile, the proper time of physical system with low kinetic energy is constant together with the space-time expansion, which suggests for accuracy to define a unit of measure (a second) based on the motion of a physical system with low kinetic energy since its proper time is invariant together with the space-time expansion.

2) Moreover in the 17th general conference on weights and measures (1983), the meter was defined as the length of the path traveled by light in a vacuum during a time interval of ${1\over c}$ of second with c=299 792 458 m/s the speed of light (\cite{WLK}). Based on this framework since the sequence $(c_n)_{n\geq0}$ decreases, then one meter will expand together with the space-time expansion and the measure of distance at successive Steps of the space-time expansion induces length contraction (if at the Step n the length $l_0= 5\ meters$ for example, then at the Step n+1 the length $l_0$ will be measured as $l_0< 5\ meters$  because the meter defined by the length of the path traveled by light in a vacuum during a time interval of ${1\over c_n}$ of second is less than the meter defined by the length of the path traveled by light in a vacuum during a time interval of ${1\over c_{n+1}}$ of second), the unit of measure changes, which affects the measure of distance.
\end{rem}

\section{Applications}

\subsection{Characteristic of Earth-Moon system since its formation}

Consider the elliptical orbit of Earth, where the farthest point (the aphelion) and the nearest point (the perihelion) from the Sun as well as the Sun lie in the $y'$-axis of the inertial reference frame $\mathcal R'_n$, such that the origin of the frame $\mathcal R'_n$ coincides with the nearest point of the orbit (Fig.\ref{Fig.0002}). An observer in the Sun rest frame $\mathcal R'_n$ reports the space  coordinates of two events that occur at the same location: the departure of the Earth from the origin and its return to it (sidereal year), as $\Delta x'_n=\Delta y'_n=\Delta z'_n=0$ and the number of full revolutions of the Moon around Earth with respect to the Sun  $\Delta t'_n=\Delta\tau_n$ as the proper time for the two events.  Then we have the following:
\begin{prop}\label{12}
 If a full revolution of the Moon around Earth with respect to the Sun (synodic month) is considered as a unit of measure of reference for time, then the proper time interval of one revolution of Earth around the Sun with respect to the positive $y'_n$-axis direction is invariant together with the space-time discrete expansion, and is equal to 12 units.
\end{prop}

{\it Proof:} Suppose that the formation of the solar system has been completed at the Step n' for $n>n'>0$. Since the average speed of the Earth around the Sun is $v=30\ km/s$, then ${v\over c_n}\simeq0.0001\in V^+(0)$, which gives that the kinetic energy of Earth is invariant together with the space-time expansion and then its proper time interval is invariant together with the space time expansion for all $n\geq n'$ (Proposition \ref{CSTPT}). Moreover if the full revolution of the Moon around Earth with respect to the Sun is considered as a unit of measure of reference for time, then following ($\S$\ref{Time}, Duration) the number of full revolutions of the Moon around Earth with respect to the Sun during one revolution of Earth around the Sun with respect to the positive $y'_n$-axis direction is determined at the Step n  (at present time) by the greatest integer value of
\begin{equation}
{\hbox{Time interval of the revolution of Earth around the Sun}\over \hbox{Time interval of a synodic month}}
\end{equation}
that is at present time
\begin{equation}
\Delta\tau_n=\Big[{\hbox{Average of one sidereal year}\over \hbox{Time interval of one Synodic month}}\Big]=\Big[{365.2564\over 29.53052}\Big]=12
\end{equation}
then
\begin{equation}
\Delta\tau_n= 12 \quad\hbox{Units},
\end{equation}
is the invariant proper time together with the space-time expansion (for all $n\geq n'$).
\begin{rem}
The Proposition \ref{12} means that the number of months in one year is 12, and the new insight is that this number has been invariant since the formation of the solar system despite the approximation of the kinetic energy of Earth (that might has a small variation due to the tidal effect). This invariance is a consequence of the fact that time dilation is imperceptible for a physical system with low level of kinetic energy (Subsection \ref{New}).
\end{rem}

\subsection{Inelastic collusion of twin particles and rest-mass}

To measure the effect of the space-time discrete expansion on energy and matter in the discrete expanding space-time $\mathcal E$, consider an inelastic collision between two identical physical systems of mass $m$ which move toward each other relative to an inertial reference frame along a straight line with equal speed of magnitude $v$. They collide and stick together relative to the same inertial reference frame in the expanding space-time $\mathcal E$. The conservation of momentum at the Step n, for all $n\geq0$, gives
\begin{equation}\label{inelas}
m_n(v)\times v-m_n(-v)\times v=M_n(V)\times V
\end{equation}
where $m_n$ is the relative mass of each particle at the step n defined by (\ref{rela}), and $M_n(V)$ is the relative mass of the new object obtained after collusion. Thus since $m_n(v)=m_n(-v)$ equality (\ref{inelas}) gives
\begin{equation}
0=M_n(V)\ V
\end{equation}
 then $V=0$. Thus, the final object is at rest with mass $M_n(0)$. Meanwhile the conservation of relative mass at the Step n, for all $n\geq0$,  gives $m_n(v) +m_n(-v)=M_n(0)$, then
\begin{equation}\label{Mn}
M_n(0)=2m \gamma_n(v),
\end{equation}
where $\gamma_n(v)$ is given by (\ref{gn}), which is the mass of the new object at rest after collision.

\begin{rem}
The mass of the final object at rest defined by (\ref{Mn}) is larger than the sum of the original masses of the twin particles at rest for each $n\geq0$. The kinetic energy of the original particles was converted into mass (\ref{Mn}) and this rest mass is affected by the discrete expansion even is $v$ is constant.
\end{rem}

\begin{prop}
Consider an inelastic collision between two identical physical systems with rest mass $m$ which move toward each other relative to an inertial reference frame along a straight line with equal speed of magnitude $v$. They collide and stick together relative to the same inertial reference frame in the expanding space-time $\mathcal E$. After collusion, the rest mass of the new object at rest is increasing together with the discrete space-time expansion and verifies for all $n>0$
\begin{equation}
M_n(0)>M_{n-1}(0).
\end{equation}
\end{prop}

{\it Proof:} Using the above inelastic collusion between two identical physical systems of mass $m$ with equal speed of magnitude $v$ along a straight line, the equality (\ref{Mn}) at the Step n and the Step n+1 (if repeated under the same conditions) gives
\begin{equation}
M_n(0)-M_{n-1}(0)=2m\Big(\gamma_n(v)-\gamma_{n-1}(v)\Big)>0
\end{equation}
since the sequence $(\gamma_n(v))_{n\geq0}$ is increasing, which concludes the proof.

\begin{rem} from the above calculus:

i) The lost kinetic energy has been converted into rest energy (mass) at each step n of the discrete space-time expansion.

ii) The classical explanation for the loss of kinetic energy attributes it to conversion into thermal energy (heat): the final object will have a higher temperature or a larger internal energy.

iii) Since the mass of the final object (after collision formula (\ref{Mn}) will have an increasing total mass together with the space-time discrete expansion), then this means that the heat of the final object will increase together with the space-time discrete expansion.

iv) If the early earth was molten from bombardment of rocks and mini planets (inelastic collusion) and others \cite{Grieve}, then based on this case study, there must exist a natural factor that contributes to the remnant heat of our planet due to the loss of kinetic energy after the inelastic collusion of rocks and mini planets attributed to the space-time expansion, a factor that maintains the remnant heat inside earth since its formation.
\end{rem}

\subsection{Nuclear fission and energy}
\subsubsection{The Binding energy}

It has been observed experimentally  that the rest mass $M(^AX)$ of a nucleus $ ^AX $ is less than the sum of the rest masses $m_i$, for all $i\in A$ of its separate pieces: protons and neutrons, (where A is the sum of the number of protons and the number of neutrons in the nucleus). The mass defect is given by
\begin{equation}
\Delta m=\sum_1^A m_i-M(^AX)
\end{equation}
and it is used to define the binding energy as the amount of mechanical energy it would take to disassemble a nucleus to its separate parts relative to a reference frame $\mathcal R_n$ at the Step n of the space-time expansion by
\begin{equation}\label{bind}
B_{{e}_n}=\Delta m\ c_n^2=\Big(\sum_1^A m_i-M(^AX)\Big)\ c_n^2.
\end{equation}

Energies such as radiation and kinetic energy are released when a nucleus undergoes a nuclear fission and splits into more stable fragments, and using the above formalism new insight can be obtained as follow.

\subsubsection{Nuclear Fission and energy liberated}

Suppose that we have an uranium nucleus $^{236}U$ in a highly excited state with nucleus rest mass $M_n(^{236}U)=M$ measured in the expanding space-time $\mathcal E$ at the step n, and suppose that the nucleus undergoes fission, at the Step n, splitting into two fragments (strontium and xenon) moving with speed $v_i$, $i=1,2$ relative to an inertial reference frame $\mathcal R_n$. It is known that these fragments rapidly emits two neutrons moving
with speed $v_3$ and $v_4$ relative to an inertial reference frame $\mathcal R_n$. The stepwise fission equation is given by
\begin{equation}
_{92}^{236}U\rightarrow _{38}^{90}Sr+\ _{54}^{142}Xe+2\ _0^1n+ Energy
\end{equation}

Thus, using the relative mass (\ref{rela}), each fragment will have a relative mass of nucleus denoted by  $m_n(^{90}Sr):=m^1_n(v_1)$, $m_n(^{142}Xe):=m^2_n(v_2)$, $m_n(neutron):=m^3_n(v_3)$, and $m_n(neutron):=m^4_n(v_4)$ at the Step n. Suppose that these pieces encounter enough material to slow them up until they stop and then each part will have a rest mass
\begin{equation}\label{restm}
m^1_n(0):=m_1,\quad m^2_n(0):=m_2,\quad m^3_n(0):=m_3,\quad \hbox{and}\quad m^4_n(0):=m_4
\end{equation}
relative to the inertial frame $\mathcal R_n$.
To reach its rest position all fragments will give an amount of energy left in the material in some forms, whatever. The left energy in the material by the fragments is denoted $K_n$ and is given by the mass-energy variation at the Step n:
\begin{equation}\label{left}
K_n=\sum_{i=1}^4 \Big(m^i_n(v_i)-m_i\Big)\ c_n^2=\sum_{i=1}^4 m_i\Big(\gamma_n(v_i)-1\Big)\ c_n^2,
\end{equation}
since using the notation (\ref{restm}) we have $m^i_n(v_i)=m^i_n(0)\gamma_n(v_i)=m_i\gamma_n(v_i)$ for $i=1,...,4$.
Moreover, the uranium nucleus has a rest mass-energy at the Step n, before the nuclear fission, given by $E_n(0)= M c_n^2$,
and after the nuclear fission the total energy of the products at the Step n is given by the sum
\begin{equation}
 m^1_n(v_1) c_n^2+m^2_n(v_2)c_n^2+m_n^3(v_3)c_n^2+m_n^4(v_4)c_n^2+E,
\end{equation}
meanwhile the conservation of total energy at the step n gives
\begin{equation}\label{conserv}
M c_n^2=\sum_{i=1}^4m^i_n(v_i) c_n^2+E
\end{equation}
if we denote the energy liberated by the nuclear fission $E:=-{\mathcal E}_{L_n}$, where the negative sign is a reference to a released energy with increase of mass by the transformation (similar notation can be found in \cite{WLK}). Thus, using (\ref{conserv}) the energy liberated  at the step n is given by
\begin{equation}\label{En2}
{\mathcal E}_{L_n}=\Big(\sum_{i=1}^4m^i_n(v_i)-M\Big) c_n^2,
\end{equation}
thus the use of (\ref{bind}) gives the binding energy
\begin{equation}\label{BE}
B_{e_n}=\Big(\sum_{i=1}^4m_i-M\Big)c_n^2
\end{equation}
 and the subtraction of (\ref{BE}) from (\ref{En2}) gives\quad ${\mathcal E}_{L_n}-B_{e_n}=K_n$, then
\begin{equation}\label{lbE}
{\mathcal E}_{L_n}=B_{e_n}+K_n
\end{equation}

The law given by (\ref{En2}) was used to estimate how much energy would be liberated under nuclear fission since the mass of uranium atom was known as well as the atoms into which it splits $ _{38}^{90}Sr$, $\ _{54}^{142}Xe$ and neutrons. It turns out from the law (\ref{En2}) that the released energy when an atom of uranium undergoes fission is affected by the space-time expansion.
\begin{cor}\label{cor4}
The sequence $(K_n)_{n\geq0}$ is increasing, with $K_n=\sum_{i=1}^4 \Big(m^i_n(v_i)-m_i\Big)\ c_n^2$ the kinetic energy of the above nuclear fission's products.
\end{cor}

{Proof:} Using the notation (\ref{restm}) we have $m^i_n(v_i)=m_i\gamma_n(v_i)$ and its substitution in the kinetic energy (\ref{left}) gives
\begin{equation}
K_n=\sum_{i=1}^4 \Big(m^i_n(v_i)-m_i\Big)\ c_n^2=\sum_{i=1}^4m_i\Big(\gamma_n(v_i)-1\Big)c_n^2=\sum_{i=1}^4m_i v_i^2\Big(f(u^i_n)\Big)
\end{equation}
with the function $f$ is given by (\ref{deri1}) and  $u^i_n={v_i\over c_n}$. The sequence $(u^i_n)_{n\geq0}$ for $i=1,..,4$ is an increasing sequence (since the sequence $(c_n)_{n\geq0}$ is decreasing and $v_i$ is a constant for $i=1,..,4$). Based on the proof of Proposition \ref{cinet}, the function $f$ is an increasing function for all $0<x<1$, then for all $n\geq0$ such that $0<u^i_n<1$ for  $i=1,..,4$, we have $u^i_{n+1}>u^i_n$ and
\begin{equation}\label{increase}
K_{n+1}-K_n=\sum_{i=1}^4m_i v_i^2\Big(f(u^i_{n+1})-f(u^i_n)\Big)>0,
\end{equation}
then for all $0<v_i<c_n$, $i=1,..,4$, we have $K_{n+1}>K_n$, which concludes the proof.

\begin{rem}
1) Based on the above corollary, the amount of energy left in the material (\ref{left}) after the nuclear fission of the nucleus is increasing together with the space-time expansion (since  $(K_n)_{n\geq0}$ is an increasing sequence), which means that the liberated heat in the surrounding material is increasing together with the space-time expansion.

2) The binding energy (\ref{BE}) of the uranium nucleus is decreasing together with the space-time expansion. Indeed, for all $n\geq0$,
\begin{equation}\label{BenDe}
B_{e_{n+1}}-B_{e_{n}}=\Big(\sum_{i=1}^4m_i-M\Big)(c_{n+1}^2-c_n^2)<0,
\end{equation}
since the sequence $(c_n)_n\geq0$ is decreasing.

2) In the above nuclear fission,  the liberated energy (\ref{lbE}) after fission  is a sum of two sorts of energies: the kinetic energy of the product (\ref{left}) due to the repulsive Coulomb force and the binding energy (\ref{BE}) that can be converted into radiation such as $\beta$ rays (electrons product), $\gamma$ rays (electromagnetic radiation). These radiations are decreasing together with the space-time expansion (see (\ref{BenDe})). This decrease in radiation is balanced with the increase of the liberated heat in the surrounding material together with the space-time expansion.
\end{rem}

\subsubsection{Natural effect on the global warming}

The origin of the Earth's heat is thought to be related to many factors but half of the Earth heat is estimated to come from radioactive decay of inside materials (\cite{Gando}). Uranium is expected to exist at the center of the Earth where it may undergo self-sustaining nuclear fission chain reactions that generate Earth major heats. Indeed, the geomagnetic field and changes in intensity are believed to be a natural consequence of variable nuclear fission chain reactions deep inside Earth (\cite{HolHern}, \cite{Hern1},\cite{Hern2},\cite{Hern3}). Moreover, it is known that nuclear fission produces energy and neutron-rich fragments that yield antineutrinos emanating, and the detection of geo-antineutrino confirmed the occurrence of nuclear fission at the center of Earth (\cite{AEF},\cite{MVJ},\cite{Domo},\cite{Fior},\cite{Gando},\cite{Mari},\cite{Rusov}) as well as the production of helium deep in earth having a ratio $^3H/^4H$ with the range observed from deep-mantle sources \cite{HolHern} that is thought to be also a consequence of deep nuclear fission, and these nuclear fission generates fifty percent of the Earth internal heat. Therefore, and based on Corollary \ref{cor4}, since the kinetic energy of the nuclear fission's fragments is increasing together with the space-time expansion, then the liberated heat in the surrounding materials in the center of Earth is increasing together with the space-time expansion, that is to say the global warming of Earth increases together with the space-time expansion if the existence of nuclear fission at the center of Earth is confirmed and this natural factor cannot be controlled.

\subsection{Galaxies missing mass problem}
\subsubsection{Doppler effect}

The doppler effect for light depends in special relativity theory (\cite{EI3}) on the relative velocity $v$ between source and observer as measured by observer, and it is expressed by the following equality
\begin{equation}\label{dop}
f_{ob}=f_{s}{{1-{v\over c}\cos\phi}\over\sqrt{{1-{v\over c}}}}
\end{equation}
where $\phi$ is the deviation of the direction of the relative velocity $v$ from the direction source-observer,  $f_{s}$ represents the proper frequency of the source and $f_{ob}$ is the frequency detected by an observer moving away with velocity $\overrightarrow{v}$ relative to the rest frame of the source. The use of the above formalism with the limiting speed $c_{n}$ given by (\ref{vl}) instead of $c$ at the Step n for all $n\geq0$ leads to the same formula for the doppler effect
\begin{equation}\label{ff0}
f_{ob}=f_{s}{{1-{v_n\over c_n}\cos\phi}\over\sqrt{{1-{v_n^2\over c_n^2}}}}
\end{equation}
where ${v}_n$ is the radial magnitude of the velocity of the observer moving away from the source as measured in the inertial reference frame $\mathcal R_n$, and $c_n$ the speed of light at the Step n for all $n\geq 0$.

\subsubsection{Spectrum line and radial velocity}

Observation of sources of light in the universe allows astronomers to use the doppler shift to determine the radial velocity of the sources as well as estimate their interior mass (\cite{SedBac}). Indeed its is known that each atom has a unique set of energy levels and it interacts with light by absorbing or emitting a specific wavelength that corresponds to the atom unique set of energy level. Each source of light in the universe is surrounded by a cloud of gas that constitute its atmosphere and when light is absorbed by atoms in the stars atmosphere, astronomers register a dark absorption line in the spectrum that corresponds to a certain kind of atom. Each atom has its own set of spectral line characteristic, thus an absorption of spectrum line allows to identify the chemical composition of the stars atmosphere under some conditions (if the star is not too cool or too hot, in intermediate temperature around $10,000\ k$ (\cite{SedBac})). The doppler effect allows to determine the motions of the stars (since light waves emitted by a moving source create a shift in color that can be measured by a spectrograph). Indeed, when the star is approaching observer, the dark absorption line in the spectrum is blue-shifted (the emitted light appears to have a shorter wavelength). When a star is moving away from observer, the dark absorption line in the spectrum is red-shifted (the emitted light appears to have a longer wavelength). The amount of change in wavelength depends on the radial velocity of the source. The radial velocity $v_n$ can be extracted from equality (\ref{ff0}). Indeed, for $\phi=0$ equality (\ref{ff0}) becomes
\begin{equation}\label{ff00}
f_{ob}=f_{s}{{1-{v_{n}\over c_n}}\over\sqrt{{1-{v^2_{n}\over c^2_n}}}}=f_{s}{{\sqrt{1-{v_n\over c_n}}}\over\sqrt{{1+{v_n\over c_n}}}},
\end{equation}
which gives the magnitude of the radial velocity $v_n$ as
\begin{equation}\label{f0}
v_{n}={f_{ob}^2-f_{s}^2\over f_{ob}^2+f_{s}^2}\ c_n.
\end{equation}

\begin{prop}\label{Cor9}
Consider a galaxy in the expanding space-time $\mathcal E$ that started emitting light at the  Step n' for $n'\ll n$ such that for all $n> n'$ an observer in an inertial reference frame at the Step n (present time) records a radial velocity for a very distant source of light given by
\begin{equation}\label{L2}
v_{n}={\lambda_{s}^2-\lambda_{ob}^2\over\lambda_{s}^2+\lambda_{ob}^2}\ c_n,
\end{equation}
and a radial velocity for a relatively distant source of light given by
\begin{equation}\label{VNL}
v_{n}={\vert{\lambda_{ob}-\lambda_{s}}\vert\over\lambda_{s}}\ c_n,
\end{equation}
then the sequence $(v_n)_{n> n'}$ for $v_n$ given by (\ref{L2}) or by (\ref{VNL}) is decreasing.
\end{prop}

{\it Proof:}
Two formula are used in astronomy, one for high speed that concerns the very distant source of light  and another for the low speed that concerns the relatively distant source of light:

i) For the very distant source of light, equation (\ref{f0}) provides the radial velocity for $f_{ob}=\di{c_n\over\lambda_{ob}}$ and $f_{s}=\di{c_n\over\lambda_{s}}$ by
\begin{equation}\label{0L2}
v_{n}={\lambda_{s}^2-\lambda_{ob}^2\over\lambda_{s}^2+\lambda_{ob}^2}\ c_n
\end{equation}
Since the chemical composition of the source atmosphere is invariant together with the space-time discrete expansion, then $\di{\lambda_{s}^2-\lambda_{ob}^2\over\lambda_{s}^2+\lambda_{ob}^2}$ is constant. Moreover $(c_n)_{n\geq0}$ is decreasing, therefore the recession radial velocity (\ref{0L2}) of the source of light verifies for all $n> n'$, $v_n > v_{n+1}$.

ii) For a relatively distant source of light that moves away with radial speed $v_{n}\ll c_n$, an approximation of equality (\ref{ff00}) neglecting terms in $\di{v_{n}^2\over c_n^2}$ and using $f_{ob}=\di{c_n\over\lambda_{ob}}$ and $f_{s}=\di{c_n\over\lambda_{s}}$ gives
\begin{equation}
f_{ob}\simeq f_{s}(1-{v_{n}\over c_n})\qquad\Longleftrightarrow\qquad{c_n\over\lambda_{ob}}\simeq{c_n\over\lambda_{s}}({1-{v_{n}\over c_n}})
\end{equation}
which gives
\begin{equation}
\lambda_{ob}\simeq \lambda_{s}({1-{v_{n}\over c_n}})^{-1}\simeq\lambda_{s}({1+{v_{n}\over c_n}}),
\end{equation}
then the radial magnitude of the speed of the source can be approximated by
\begin{equation}\label{0VNL}
v_{n}={\vert\Delta \lambda\vert\over\lambda_{s}}c_n={\vert{\lambda_{ob}-\lambda_{s}}\vert\over\lambda_{s}}\ c_n.
\end{equation}
where $\Delta \lambda$ is positive for the red-shift  and $\Delta \lambda$ is negative for the blue-shift.
 Since the chemical composition of the source atmosphere is invariant together with the space-time expansion, then  ${\vert\Delta \lambda\vert\over\lambda_{s}}$ is constant. Moreover the sequence  $(c_n)_{n\geq0}$ is decreasing, therefore the radial magnitude of the speed of the source (\ref{0VNL}) verifies for all $n> n'$, $v_{n}>v_{n+1}$.

\begin{rem}
1) In both cases of the above proof, we find that the radial velocity is decreasing together with the space-time discrete expansion, which suggests that the source's radial velocity was greater in the past, and this output has an enormous impact in astronomy. Indeed a small variation in the speed of light causes an enormous error in distance measurement to stars as well as in galaxies and planets' mass estimation.

2) The radial sequence $(v_n)_{n> n'}$ has the same monotony as the sequence $(c_n)_{n> n'}$ where at the Step n' the galaxy started emitting light. Indeed based on Property \ref{light}, the sequence  $(v_n)_{n> n'}$ is then constant for a small scale of time interval and variable for a large scale of time interval. It means that an observer records at present time a constant value of the radial velocity during a short scale of time period, however if the observer records the radial velocity during a large scale of time interval, he will find it decreasing.

3) Due to the finite limiting speed of light (\ref{vl}), observing distant galaxies conveys a look back in time that corresponds to the time (in years) needed by the light to travel from the distant galaxy to reach Earth. The more distant galaxy you are looking at, the larger scale of time you look back to the galaxy as it was at a time long before when the universe was knowingly different. Then looking back in time by observation suggests (based on Proposition \ref{Cor9}) that the dynamic of rotation of all galaxies and its constituents was faster in the past (including our galaxy), the dynamic of galaxies and its constituents decreases together with the space-time expansion. This result concords with the outcomes in Paleochronobiology  (study of organisms that record the resulting changing geophysical periodicity in their skeletons) establishing evidence for the fact that earth rotation was faster in the past (\cite{Wells},\cite{Rosen},\cite{Pannella}).
\end{rem}

\subsubsection{The galaxies missing mass problem}

The galaxies rotation problem consists of the discrepancy between the observed rotation of galaxy and the newtonian gravity prediction. This discrepancy led to the consideration of the existence of an invisible matter (\cite{perc}) to compensate the missing mass. The property of the radial velocity of the  light's source relative to an observer at the Step n has a major consequence on the missing mass for galaxies. Indeed almost all measurements of motion in astronomy use the doppler shift effect. The mass of galaxies are then calculated via the determination of the radial velocity of the  light's source relative to an observer using spectrograph (\cite{RuFord}, \cite{Rub}, \cite{RubTh}) of stars distribution in galaxies. When astronomers observe distant galaxies at present time  (Step n), they observe galaxies as they were in the past when they started emitting light at a Step n' with $n'\ll n$ (from the past), and the light reaches the observer at the Step n (present time). Thus estimation of the galaxies mass using a constant radial velocity of the  light's source  will be erroneous if the radial velocity of the light's source was greater in the past (Proposition \ref{Cor9}).
\begin{prop}
Consider a galaxy in the space-time $\mathcal E$ that started emitting light at the Step n', for $n'\ll n$ and a star with mass $m_n$ orbiting the core of the galaxy at a radius r , such that an observer in an inertial reference frame at the step n (present time) records its radial velocity $v_{n}$ for all $n> n'$ given by (\ref{L2}) or by (\ref{VNL}). Then the enclosed mass to the orbital radius r of the star is given by $M_n=r{v_{n}^2\over G}$, and the sequence $(M_n)_{n>n'}$ is decreasing together with the space-time expansion.
\end{prop}

{\it Proof:} Consider a star orbiting the center of a given galaxy in the space-time $\mathcal E$ with discrete expansion, such that the star experiences a gravitational force due to the galaxy mass given by $F={GM_nm_n\over r^2}$, where $M_n$ is the galaxy mass interior to the orbital radius $r$ and  $m_n$ is the star mass as measured by an observer in an inertial reference frame $\mathcal R_n$ at the Step n. The star in its circular orbit must experience a centripetal acceleration of magnitude $\gamma={v_n^2\over r}$ and the Newton second law gives $F=m_n\gamma$, that is to say ${GM_nm_n\over r^2}=m_n{v^2\over r}$, which provides the mass $M_n={v_n^2r\over G}$ where $v_{n}$ is given by (\ref{VNL}) for a relatively distant galaxy, or by (\ref{L2}) for a distant galaxy. Since the sequence $(v_{n})_{n>n'}$ is decreasing together with the space-time expansion (Proposition \ref{Cor9}), then the inner measured mass of the galaxy in the reference frame $\mathcal R_n$ at the Step n (the present time) verifies
\begin{equation}
M_n={v_n^2r\over G}<{v_{n-1}^2r\over G}=M_{n-1}
\end{equation}
that is to say for all $n>n'$, $M_n<M_{n-1}$ which complete the proof.
\begin{rem}
The above result suggests that the measured galaxy mass, interior to the orbital radius of the star, at the Step n is smaller than the mass of the galaxy at the Step n' for $n>n'$, that is to say the approximation of the mass of the galaxy at present time decreases together with the space-time expansion. Thus the difference $\Delta M=M_{n'}-M_{n}>0$ might be the real cause of the discrepancy between observation and theory that creates the problem of galaxies missing mass and its interpretation.
\end{rem}

\section{Conclusion}
Within this case study, the Lorentz transformations obtained (\ref{E4g}) in the Proposition \ref{Ltransf} verify all the properties provided by the special relativity formalism, including the invariance of the limiting velocity, the non simultaneity of events, the relative time dilation, and the length contraction. This extension takes into account consequences of a small variation of the limiting velocity for a large scale of time together with the space-time expansion and provides interesting approach and interpretation of unexplored phenomena in physics such as:

\begin{itemize}
  \item discovery of a natural boundary (barrier) between physical systems with high and low kinetic energy.

  \item different characteristics of the proper time for physical system with high kinetic energy and physical system with low kinetic energy together with the space-time expansion.

      \item  a new insight for Earth revolution around the Sun since its formation, and a new factor that increases Earth's earlier remnant inside heat as well as the Earth's inner heat due to radioactive decay or nuclear fission of inside materials. This might revise our understanding of the factors effect for the  Earth's global warming problem

  \item the decrease of the binding energy of matter together with the space-time expansion and the increase of the liberated heat in the surrounding material together with the space-time expansion for nuclear fission. The decrease of the binding energy together with the space-time expansion might lead to revise the estimation period of radiation of the nuclear waste problem.

  \item a new lead to explain the discrepancy between observation and theory that creates the problem of galaxies missing mass.
\end{itemize}
 This case study led to build an adequate definition of time and its characteristics together with the space-time expansion.

\bibliographystyle{spphys}       

\end{document}